\documentstyle[12pt]{article}

\def\anti{\buildrel \wedge \over }
\def\symm{\buildrel \sim \over }
\def\ps{{}^{\mbox{\tiny p}}\!}
\def\vect{\buildrel \rightharpoonup \over }
\def\tens{{}^{\mbox{\tiny T}}\!\!}

\def\a{\alpha }
\def\be{\beta }

\def\g{\gamma }
\def\d{{\cal D}}
\def\de{\delta }

\def\disp{\displaystyle}
\def\e{\epsilon }

\def\fl{{\flat}}
\def\h{{\cal H}}
\def\ve{\varepsilon }
\def\th{e}

\def\k{\kappa }
\def\la{\lambda }
\def\La{\Lambda }
\def\lg{{\cal L}}
\def\lr{\longrightarrow }
\def\lan{\langle }
\def\na{\nabla \! }

\def\G{{\mit\Gamma}}
\def\p{\partial }
\def\ol{\overline }
\def\ran{\rangle }

\def\rmc{{\mbox{\scriptsize can}}}
\def\rmd{{\mbox{d}}}

\def\rmT{{\mbox{\scriptsize T}}}
\def\s{\sigma }

\def\ul{\underline }

\def\bc{\bigcirc}
\def\ob{\overbrace}

\catcode`\@=11
\long\def\@makefntext#1{
\protect\noindent \hbox to 3.2pt {\hskip-.9pt
$^{{\eightrm\@thefnmark}}$\hfil}#1\hfill}               

\def\@makefnmark{\hbox to 0pt{$^{\@thefnmark}$\hss}}    

\def\ps@myheadings{\let\@mkboth\@gobbletwo
\def\@oddhead{\hbox{}
\rightmark\hfil\eightrm\thepage}
\def\@oddfoot{}\def\@evenhead{\eightrm\thepage\hfil
\leftmark\hbox{}}\def\@evenfoot{}
\def\sectionmark##1{}\def\subsectionmark##1{}}

\newcommand{\smalllineskip}{\baselineskip=10pt}
\renewenvironment{thebibliography}[1]
        {\frenchspacing
          \baselineskip=18pt
         \begin{list}{\arabic{enumi}.}
        {\usecounter{enumi}\setlength{\parsep}{0pt}
         \setlength{\leftmargin 12.7pt}{\rightmargin 0pt} 
         \setlength{\itemsep}{0pt} \settowidth
        {\labelwidth}{#1.}\sloppy}}{\end{list}}

\newcommand{\fcaption}[1]{
        \refstepcounter{figure}
        \setbox\@tempboxa = \hbox{\footnotesize Fig.~\thefigure. #1}
        \ifdim \wd\@tempboxa > 5in
           {\begin{center}
        \parbox{5in}{\footnotesize\smalllineskip Fig.~\thefigure. #1}
            \end{center}}
        \else
             {\begin{center}
             {\footnotesize Fig.~\thefigure. #1}
              \end{center}}
        \fi}

\newcommand{\tcaption}[1]{
        \refstepcounter{table}
        \setbox\@tempboxa = \hbox{\footnotesize Table~\thetable. #1}
        \ifdim \wd\@tempboxa > 5in
           {\begin{center}
        \parbox{5in}{\footnotesize\smalllineskip Table~\thetable. #1}
            \end{center}}
        \else
             {\begin{center}
             {\footnotesize Table~\thetable. #1}
              \end{center}}
        \fi}

\def\@citex[#1]#2{\if@filesw\immediate\write\@auxout
        {\string\citation{#2}}\fi
\def\@citea{}\@cite{\@for\@citeb:=#2\do
        {\@citea\def\@citea{,}\@ifundefined
        {b@\@citeb}{{\bf ?}\@warning
        {Citation `\@citeb' on page \thepage \space undefined}}
        {\csname b@\@citeb\endcsname}}}{#1}}

\newif\if@cghi
\def\cite{\@cghitrue\@ifnextchar [{\@tempswatrue
        \@citex}{\@tempswafalse\@citex[]}}
\def\citelow{\@cghifalse\@ifnextchar [{\@tempswatrue
        \@citex}{\@tempswafalse\@citex[]}}
\def\@cite#1#2{{$\null^{#1}$\if@tempswa\typeout
        {IJCGA warning: optional citation argument
        ignored: `#2'} \fi}}

\def\pmb#1{\setbox0=\hbox{#1}
        \kern-.025em\copy0\kern-\wd0
        \kern.05em\copy0\kern-\wd0
        \kern-.025em\raise.0433em\box0}

\font\eightrm=cmr8

\def\qed{\hbox{${\vcenter{\vbox{                        
   \hrule height 0.4pt\hbox{\vrule width 0.4pt height 6pt
   \kern5pt\vrule width 0.4pt}\hrule height 0.4pt}}}$}}

\title{\Large\bf HAMILTONIAN ANALYSIS OF POINCAR\'E GAUGE THEORY:
HIGHER SPIN MODES}

\author{\large HWEI-JANG YO\thanks{E-mail address: hjyo@asiaa.sinica.edu.tw}\\
\large\it Institute of Astronomy and Astrophysics,\\
\large\it Academia Sinica, Taipei 115, Taiwan, ROC\\
\large\it also Department of Physics, University of Illinois\\
\large\it at Urbana-Champaign, Urbana, Illinois 61801, USA\\
\ \\
\large JAMES M. NESTER\thanks{E-mail address: nester@phy.ncu.edu.tw}\\
\large\it Department of Physics,\\
\large\it National Central University, Chungli 320, Taiwan, ROC Fy; 04.50 +h\hspace*{\fill}\\
\rm\normalsize
Short title: {\bf Hamiltonian Analysis of PGT: Higher spin modes}
\hspace*{\fill}}
\date{\today}

\begin{document}
\maketitle
\begin{abstract}
We examine several higher spin modes of the
Poincar\'e gauge theory (PGT) of gravity using the Hamiltonian analysis.
The appearance of certain undesirable effects due to non-linear
constraints in the Hamiltonian analysis
are used as a test.  We find that
the phenomena of field activation and constraint
bifurcation both exist in the pure spin 1 and the pure spin 2 modes.
The coupled spin-$0^-$ and spin-$2^-$ modes also fail our test due to
the appearance of constraint bifurcation.
The ``promising'' case in the linearized theory of PGT given by Kuhfuss
and Nitsch\cite{KRNJ86} likewise does not pass.
From this analysis of these specific PGT modes we conclude that an
examination of such nonlinear constraint effects shows great promise as
a strong test for this and other alternate theories of gravity.
\end{abstract}

\section{Introduction}
\noindent
Many alternative theories of gravity have been proposed to replace
general relativity (GR);
experimental and observational results are applied to examine the
predictions of each alternative and to eliminate the unsatisfactory
ones\cite{WilC81}.
Certain gauge theories of gravity, with their many ranges of parameters,
not only pass all the experimental and observational tests but also
agree with GR to Post-Newtonian order\cite{SMSN79}.
Consequently no discriminating experimental and observational test
can presently distinguish them from GR.
Thus they are observationally equally viable.
Spontaneously many people paid attention to another criteria ---
theoretical tests.
Essentially, besides self-consistency, theoretical tests judge if
a theory obeys commonly accepted physical requirements, such as
``no tachyons'' (faster than light signals) and ``no ghosts'' (negative
energy waves).
Moreover a viable theory should have an appropriate mathematical structure.
This includes a well-posed initial value problem:  the basic requirement
is the Cauchy-Kovalevska theorem; beyond that, the propagation of
dynamic modes should be described by hyperbolic evolution equations
with well-behaved characteristics.

Although the metric-affine gauge theory of gravity (MAG)\cite{HMMN95}
is more general (it has curvature, torsion and non-metricity), we here
apply our Hamiltonian based analysis mainly
to the Poincar\'e gauge theory (PGT), an important sub-theory (with only
curvature and torsion, i.e.,  vanishing nonmetricity) of the MAG.
One reason for this choice is physical.
In the foreseeable future the effect of nonmetricity is thought to be
extremely hard to measure as it is expected to only survive at
the Planck energy scale.
On the other hand it is believed that the effect of torsion has the
possibility to be detected from some strong astronomical sources.
However our primary motivation for restricting our considerations to the
PGT is that it is simpler. Before examining nonlinear effects in the MAG,
we really need more experience.
Also there have been many more studies of
the behavior of the PGT.  These works provide a support
for our deeper investigation.

Pioneered by Sezgin and Nieuwenhuizen\cite{SENP80}, people examined
the PGT based on the ``spin-projection operator formalism''\cite{NieP73}
and the weak field approximation with positive energy
and mass\cite{HKST80}.
They put restrictions on the ranges of the PGT parameters and gave
tables of ``viable''  PGT parameter ranges with well-behaved propagating
modes. The positive energy test (PET),
which excludes the cases permitting a non-trivial
solution that has non-positive energy, was proposed as an effective
theoretical test for gravitational theories and applied to
the PGT\cite{CDNJ86,CDJY92}.
Parameter restrictions leading to the interesting purely curvature
or torsion modes were identified\cite{KatM93}.
On the other hand, Cauchy-Kovalevska and hyperbolicity conditions
were imposed on the PGT to study the initial value problems\cite{DimA89};
a shock wave analysis was applied to find the characteristics of
propagating modes\cite{LemJ90}.
However, it should be noted that nonlinear effects were essentially
not probed in most of these investigations.
Almost all of the above examinations were directly or effectively working
on the linearization of the PGT, except for the PET.

The PET has been applied to the PGT and gave some good results with only
simple ansatz which are used to find ``bad'' solutions\cite{YoHJ91}.
The nice feature is that the PET considers the full nonlinear theory
and thus is not subject to such doubts that a result of a test derived
from the linearized theory suffers.
However, the test can be applied only if the total energy is well defined,
hence it applies only to the asymptotically flat or constant negative
curvature configurations.
Actually, in practice the PET has not given many new results beyond
those found from the linearized theory tests, simply because it is not so
easy to apply in general.
This limitation provided further motivation for our nonlinear
Hamiltonian investigation.

An early warning of the difficulties (due to nonlinearity) was raised by
Kopczy\'nski\cite{KopW82}.
He noted that the one parameter teleparallel theory
(OPTP, a.k.a.${}$ NGR)\cite{HKST79} had predictability problems.
Further analysis\cite{NesJ88} revealed that the problem
was not generic, it occurred only for a special class of solutions.
Cheng {\it et al} using a Hamiltonian approach
verified this property\cite{CWCN88} and identified
the troubles with the phenomenon of {\it constraint bifurcation}:
the chain of constraints could ``bifurcate'' depending on the values
of the fields, so that the number and/or type of constraints would
depend on the values of the phase space variables.
Since the OPTP is a sub-theory (with curvature vanishing) of the PGT,
it was expected that the PGT as well as the MAG and other alternate
geometric gravity theories were also vulnerable to this difficulty.
Besides constraint bifurcation, it is likely that the gauge theories of
gravity will encounter the nonlinear phenomenon of turning constraints
into equations of motion as in the analysis by Velo and Zwanzinger\cite{Vel269}
of higher-spin fields.
Also within Einstein's theory of GR with sources included,
the inherent non-linearity of the gravitational coupling to
sources has long been known to have peculiar effects, including
the activation of non-dynamic source
degrees of freedom (see, in particular, Ref.~19 
and \S 5.2 in Ref.~20). 
These interesting considerations motivated our investigation
which is reported in this paper.

Our main tool for studying these nonlinear effects is the Hamiltonian
analysis using the Dirac constraint algorithm in the full nonlinear PGT.
The Hamiltonian analysis reveals whether a theory is
consistent.
It systematically determines the constraints, the gauges, and the degrees
of freedom for each consistent theory.
The analysis also prescribes methods for reducing the phase space of
the theory to one containing only the degrees of freedom\cite{IJNJ80}.
Moreover it is the most straightforward way to find the apparatus of
the initial value scheme for a given theory --- the proper data,
the constraints, and the evolution equations.

We base our test of the viability of a particular case of the PGT (with
certain parameter choices)
by checking if the whole structure of the theory stays the same before and
after linearization.
The change of the structure of the theory can be measured by
studying the nonlinearity in the Poisson matrix of the constraints.
In practice this was not so complicated, for in all the cases we have
examined it was sufficient to consider the Poisson matrix of the
{\em primary} constraints.
In Ref.~21 
we analyzed two modes of PGT with propagating spin-$0$ torsion.
In this paper we continue our examination of the PGT considering now the
simple spin-$1$
modes, simple spin-$2$ modes, coupled spin-$0^-$ and spin-$2^-$ modes, and
the modes of the preferred parameter choice identified by
Kuhfuss and Nitsch\cite{KRNJ86}.
We find that two phenomena, i.e., {\it constraint bifurcation} and
{\it field activation}, could happen in all of these cases\cite{YoHJ99}.
These phenomena prevent these cases from being viable.

This paper is organized as follows.  In section 2 
we review the
basic elements of the PGT and introduce its Lagrangian and field equations.
In section 3 
we use the Dirac theory for constrained Hamiltonian
systems in the neat ``if'' constraint formulation developed by
Blagojevi\'c and Nikoli\'c\cite{BMNI83,NikI84}.  The primary
constraints, including ten ``sure'' primary constraints and thirty
so-called ``primary if-constraints'', are given.  The total
Hamiltonian density, including the canonical Hamiltonian density
and all possible primary constraints, is presented.
We summarize the results and the viable parameter combinations of the
linearized PGT in section 4. 
The viable parameter combinations in this
section are a guide to search for the viable modes of the full PGT.
In section 5 
we study the viability of the PGT of the
simple spin-$1$ modes, simple spin-$2$ modes, coupled spin-$0^-$
and spin-$2^-$ modes, and give an argument to explain the failure
of the parameter choice by Kuhfuss and Nitsch.
In the final section we present our conclusions.

Throughout the paper our conventions are basically the same as Hehl's
in Ref.~25. 
We have made a few adjustments to accommodate the translation
of the Hamiltonian ``if'' constraint formalism to these conventions.
The Latin indices are coordinate (holonomic) indices, whereas the
Greek indices are orthonormal frame (an-holonomic) indices.
The first letters of both alphabets ($a, b, c, \dots$; $\a, \be, \g,
\dots$) run over 1, 2, 3, whereas the later ones run over 0, 1, 2, 3.
Furthermore, $\eta _{\mu \nu}=$diag($-$, $+$, $+$, $+$); $\e ^{\mu \nu \g \de}$
is the completely antisymmetric tensor with $\e ^{{\hat 0}{\hat 1}
{\hat 2}{\hat 3}}=-1$.
The meaning of a bar over a Greek index is adopted from Ref.~24.

\section{Poincar\'e Gauge Theory of Gravitation}
\label{sec2}
\noindent
In the PGT there are two sets of gauge potentials, the orthonormal
 frame field (tetrads) $\th _i{}^\mu$ and the metric-compatible connection
 $\G _{i\mu}{}^\nu$, which are associated with the translation and the
Lorentz
 subgroups of the Poincar\'e gauge group, respectively. The associated
 field strengths are the torsion
 \begin{equation}
  T_{ij}{}^\mu =2(\p _{[i}\th _{j]}{}^\mu
  +\G _{[i|\nu }{}^\mu \th _{|j]}{}^\nu),
 \end{equation}
 and the curvature
 \begin{equation}
  R_{ij\mu}{}^\nu =2(\p _{[i}\G _{j]\mu}{}^\nu
  +\G _{[i|\s }{}^\nu \G _{|j]\mu}{}^\s ),
 \end{equation}
 which satisfy the  Bianchi identities
 \begin{eqnarray}
  &&\na _{[i}T_{jk]}{}^\mu \equiv R_{[ijk]}{}^\mu,\\
  &&\na _{[i}R_{jk]}{}^{\mu \nu}\equiv 0.
 \end{eqnarray}
 The conventional form of the action, which is invariant under the
 Poincar\'e gauge group, has the form
 \begin{equation}
  S=\int d^4x\th(L_M+L_G),
 \end{equation}
 where $L_M$ stands for the matter Lagrangian density (which
determines the energy-momentum and spin source currents),
$L_G$ denotes
the gravitational Lagrangian density, and $\th =det(\th _i{}^\mu )$.
In this paper we are concerned with the gravitational propagating
modes,  hence we omit
the matter Lagrangian density, so $L_G$ is considered
as the source-free total Lagrangian. Varying
with respect to the potentials then gives the (vacuum) field equations,
 \begin{eqnarray}
  &&\na _jH_\mu {}^{ij}-\ve _\mu {}^i=0,\\
  &&\na _jH_{\mu \nu}{}^{ij}-\ve _{\mu \nu}{}^i=0,
 \end{eqnarray}
 with the field momenta
 \begin{eqnarray}
  H_\mu {}^{ij}&:=&{\p \th L_G\over \p \p _j\th _i{}^\mu}
  =2{\p \th L_G\over \p T_{ji}{}^\mu},\\
  H_{\mu \nu}{}^{ij}&:=&{\p \th L_G\over \p \p _j\G _i{}^{\mu \nu}}
  =2{\p \th L_G\over \p R_{ji}{}^{\mu \nu}},
 \end{eqnarray}
 and
 \begin{eqnarray}
  \ve _\mu {}^i&:=&e^i{}_\mu \th L_G-T_{\mu j}{}^\nu H_\nu {}^{ji}
  -R_{\mu j}{}^{\nu \s}H_{\nu \s}{}^{ji},\\
  \ve _{\mu \nu}{}^i&:=&H_{[\nu \mu]}{}^i.
 \end{eqnarray}
The Lagrangian is chosen (as usual) to
be at most of quadratic order in the field strengths,
then the field momenta are linear in the field strengths:
 \begin{eqnarray}
  H_\mu {}^{ij}&=&{\th \over l^2}\sum^3_{k=1} a_k
  {\buildrel {(k)} \over T}{}^{ji}{}_\mu,\\
  H_{\mu \nu}{}^{ij}&=&-{a_0\th \over l^2}e^i{}_{[\mu}e^j{}_{\nu ]}
  +{\th \over \k}\sum^6_{k=1}b_k
  {\buildrel {(k)} \over R}{}^{ji}{}_{\mu \nu},
 \end{eqnarray}
the three $\displaystyle {{\buildrel {(k)} \over T}{}^{ji}{}_\mu}$
and the six $\displaystyle { {\buildrel {(k)} \over R}{}^{ji}{}_{\mu
\nu}}$ are the algebraically irreducible parts of the torsion and the
curvature, respectively. The reciprocal frames $e^i{}_\mu$ and
$\th _i{}^\mu$ satisfy $e^i{}_\mu \th _i{}^\nu =\de _\mu {}^\nu$
and $e^i{}_\mu \th _j{}^\mu =\de _j{}^i$; the coordinate metric
is defined by $g_{ij}=\th _i{}^\mu \th _j{}^\nu \eta _{\mu \nu}$.
$l$ and $\k$ are constants;
the $a_k$ and $b_k$ are free coupling parameters. Due to
the Bach-Lanczos identity only five of the six $b_k$'s are
independent.  $a_0$ is the coupling parameter of the scalar curvature
$R:=R_{\mu \nu}{}^{\nu \mu}$.
 For the Hamiltonian
formulation we associate the {\it canonical momenta} with certain components
of the covariant field momenta:
 \begin{equation}
  \pi^i{}_\mu \equiv H_\mu {}^{i0},\quad
  \pi^i{}_{\mu \nu}\equiv H_{\mu \nu}{}^{i0}.\label{tm}
 \end{equation}

\section{Total Hamiltonian and Primary If-constrains}
\label{chap5}
\noindent
The Hamiltonian analysis is our main tool for studying the PGT.
We present here the primary constraints and the total Hamiltonian density
in terms of the decomposition of the canonical variables.
As indicated in Table \ref{cpc}, in addition to certain sure constraints
associated with the basic local Poincar\'e gauge symmetry of spacetime,
for each of the kinetic critical parameter combinations which vanishes
there is an extra degeneracy.
Such extra degeneracies are associated with {\it extra} primary constraints
which can lead to secondaries etc.
Essentially we follow the techniques developed by Blagojevi\'c
and Nikoli\'c,
who investigated systematically all of these possibilities of the
critical parameter combinations\cite{BMNI83,NikI84}.
They adopted the {\it if-constraint} technique which avoids getting stuck in
unnecessary detail and still identifies all possible constraints.

The sure primary constraints are
\begin{equation}\label{spc}
      \pi^0{}_\mu\approx0,\quad
      \pi^0{}_{\mu\nu}\approx0.
\end{equation}
These constraints reflect the fact that the torsion and the curvature are
defined as the antisymmetric derivatives of $\th _i{}^\mu$ and $\G _i{}^{\mu
\nu}$; they do not involve the ``velocities'' ${\dot\th}_0{}^\mu$ and ${\dot
\G}_0{}^{\mu \nu}$.
One will obtain further primary ``if''-constraints if the
Lagrangian density is singular with respect to any parts of
the remaining ``velocities'',
$\dot\th _a{}^\mu$, and $\dot\G _a{}^{\mu \nu}$.
The total Hamiltonian density is formally of the form
\begin{equation}
   \h_\rmT=\h_\rmc +u_0{}^\mu \pi ^0{}_\mu
    +{1\over 2}u_0{}^{\mu \nu}\pi ^0{}_{\mu \nu}+u^A\phi _A,
\label{thd}
\end{equation}
where the $\phi _A$ are the primary ``if'' constraints and the $u$'s denote the
associated {\it Lagrange multipliers}; $\h_\rmc$ stands for the canonical
Hamiltonian density which will be specified below.
The primary constraints (\ref{spc}) are first-class, i.e., $\pi ^0{}_\mu$
and $\pi ^0{}_{\mu \nu}$ are unphysical variables. By using the Hamilton
equation of motion we can infer that the multipliers $u_0{}^\mu$ and
$u_0{}^{\mu\nu}$ are indeed equal to $\dot \th _0{}^\mu$ and $\dot \G _0{}^{\mu
\nu}$.  They are dynamically undetermined pure gauge multipliers.

In the PGT we consider a nine-parameter Lagrangian  which is of
the $R+T^2+R^2$ type.
It is well-known that the ``most dynamical'' case of the theory, when the
parameter combinations are not critical, is of no physical importance due to
the presence of ghosts and tachyons\cite{HKST80}, one will expect there are
``{\it primary if-constraints}''(PIC).
Sec.~3.2 
is devoted to the subject.
Here we outline the procedure as follows:
The first step is to find all the values of parameters which diminish
the rank of the Hessian matrices
$\p^2\lg/\p{\dot\th}_a{}^\mu\p{\dot\th}_b{}^\nu$ and
$\p^2\lg/\p{\dot\G}_a{}^{\mu\nu}\p{\dot\G}_b{}^{\tau\s}$ --- the critical
values of the parameters.
Such values of parameters result in some torsion modes being frozen.
We will show all of the possible primary constraints,
which appear when the parameters take on the critical values.
They will be written in the form of ``if''-constraints, which automatically
drop out from the theory when the corresponding critical values of the parameters
are not fulfilled.
Finally, we will find the expression for the super-Hamiltonian
$\h_\bot$, which is valid for all values of the parameters.

\subsection{Total Hamiltonian}
\noindent
Before we proceed to obtain the explicit form of the canonical
Hamiltonian density and $\phi _A$,
it is useful to define the decomposition of the related variables and functions.
The components of the unit normal {\bf n} to the $x^0=$constant hypersurface,
with respect to the orthonormal frame, are given by
\begin{equation}
   n_\mu :={-e^0{}_\mu \over \sqrt {-g^{00}}}.
\end{equation}
A vector, e.g., $V_\mu$, can be decomposed into the orthogonal and parallel
components with respect to the orthonormal frame indices:
\begin{eqnarray}
   V_\mu &=&-V_\bot n_\mu +V_{\ol \mu },\\
   V_\bot &\equiv &V_\mu n^\mu,\\
   V_{\ol {\mu}}&\equiv &V_\nu(\de _\mu {}^\nu +n_\mu n^\nu ).
\end{eqnarray}
One can extend the decomposition to any tensors with orthonormal frame indices.
The lapse and shift functions can be written as
\begin{eqnarray}
   N&\equiv &{1\over \sqrt {-g^{00}}}=-n_\mu \th _0{}^\mu ,\\
   N^a&\equiv &-{g^{0a}\over g^{00}}=\th _0{}^\mu e_{\ol \mu }{}^a,
\end{eqnarray}
and $\th =-NJ$,
where $(-J)$ is the determinant of the 3-metric.

To construct the canonical Hamiltonian density, we decompose the torsion and
the curvature tensor as
\begin{eqnarray}
   T_{\nu\s}{}^\mu&=&2n_{[\nu}T_{\ol\s]\bot}{}^\mu+T_{\ol{\nu\s}}{}^\mu,\\
   R_{\tau\s}{}^{\mu\nu}&=&2n_{[\tau}R_{\ol\s]\bot}{}^{\mu\nu}+R_{\ol
    {\tau\s}}{}^{\mu\nu},
\end{eqnarray}
so that $T_{\ol{\nu\s}}{}^\mu$ and $R_{\ol{\tau\s}}{}^{\mu\nu}$ are independent
of velocities.
Defining the convenient ``parallel'' canonical momenta
\begin{equation}
   \pi ^{\ol \s}{}_\mu \equiv \th _a{}^\s \pi ^a{}_\mu,\quad
   \pi ^{\ol \s}{}_{\mu \nu}\equiv \th _a{}^\s \pi ^a {}_{\mu \nu},
\end{equation}
which satisfy $\pi ^{\ol \s}{}_\mu n_\s =0$, $\pi ^{\ol \s}{}_{\mu \nu}n_\s =0$,
we obtain
\begin{equation}\label{mou1}
      \pi^{\ol\s}{}_\mu\equiv\th_a{}^\s\disp{\p\th\lg\over\p{\dot\th}_a{}^\mu}
       =J\disp{\p\lg\over\p T_{\ol\s\bot}{}^\mu},\quad
      \pi^{\ol\s}{}_{\mu\nu}\equiv\th_a{}^\s\disp{\p\th\lg\over\p{\dot\G}
       _a{}^{\mu\nu}} =J\disp{\p\lg\over\p R_{\ol\s\bot}{}^{\mu\nu}}.
\end{equation}
The canonical Hamiltonian density,
\begin{equation}
   \h_\rmc=\pi ^a{}_\mu {\dot \th}_a{}^\mu +{1\over 2}\pi ^a{}
    _{\mu \nu}{\dot \G}_a{}^{\mu \nu}-\th\lg,
\end{equation}
can be rewritten in the so-called Dirac-ADM form\cite{DirP64,HART76,SunK82},
\begin{equation}
   \h_\rmc=N\h _\bot +N^a\h _a +{1\over 2}\G_0{}^{\mu \nu}\h _{\mu \nu}
    +\p _a\d ^a,\label{chd}
\end{equation}
which is linear in $N$ and $N^a$.
The other quantities are given by\cite{NikI84}
\begin{eqnarray}
   \h_\bot&=&\pi ^{\ol \s}{}_\mu T_{\bot \ol \s}{}^
    \mu +{1\over 2}\pi ^{\ol \s}{}_{\mu \nu}R_{\bot \ol \s}
    {}^{\mu \nu}+J\lg-n^\mu \na _a\pi ^a{}_\mu,\label{ht}\\
   \h _a&=&\pi ^b{}_\mu T_{ab}{}^\mu +{1\over 2}\pi ^b{}_
    {\mu \nu}R_{ab}{}^{\mu \nu}-\th _a{}^\mu \na _b
    \pi ^b{}_\mu,\label{superm} \\
   \h _{\mu \nu}&=&\pi ^a{}_\mu \th _{a\nu}-\pi ^a{}_\nu
    \th _{a\mu}-\na _a\pi ^a{}_{\mu \nu},\label{lorentzp}\\
   \d^a&=&\pi^a{}_\mu\th_0{}^\mu+{1\over2}\pi^a{}_{\mu\nu}\G_0{}^{\mu\nu}.
\end{eqnarray}
With the help of Eq.~(\ref{mou1}) we can rewrite the expression in the
parentheses in Eq.~(\ref{ht}) as a function of $T_{\ol{\nu\s}}{}^\mu$,
$\pi^{\ol\s}{}_\mu$, $R_{\ol{\tau\s}}{}^{\mu\nu}$ and $\pi^{\ol\s}{}_{\mu\nu}$.
Thus it is independent of the unphysical variables and represents the only
dynamical part of the canonical Hamiltonian density.
By combining (\ref{thd}) and (\ref{chd}) the total Hamiltonian is
\begin{eqnarray} \label{thf}
   \h_\rmT&=&N\h_\bot+N^a\h_a+{1\over2}\G_0{}^{\mu\nu}\h_{\mu\nu}+\p_a\d^a
    \nonumber\\
   &&+u_0{}^\mu\pi^0{}_\mu+{1\over2}u_0{}^{\mu\nu}\pi^0{}_{\mu\nu}+u^A\phi_A.
\end{eqnarray}
According to Dirac's general arguments, there have to be (at least) ten
first-class constraints in the PGT,  as it is invariant under the
ten-parameter local Poincar\'e group.
By utilizing the  total Hamiltonian density (\ref{thf}) in the consistency
conditions for the sure primary constraints (\ref{spc}) we obtain the
{\it sure secondary constraints} (SC)
\begin{equation}
   \h_\bot\approx0,\quad\h_a\approx0,\quad\h_{\mu\nu}\approx0.
\end{equation}
$\h_\bot$ and $\h_a$ represent the generators of the orthogonal and
parallel $x^0=$const hypersurface deformations\cite{ARDM62,TeiC73,KucK74}.
And $\h_{\mu\nu}$ represents the generators of the local Lorentz
transformations\cite{DirP62,DSIC76,NJTC78}.
They are all first-class constraints, so that their consistency conditions
are trivially satisfied\cite{CasL82}.
Only the super-Hamiltonian $\h _\bot$ is involved in the dynamical evolution,
the super-momenta $\h _a$ and Lorentz rotation parts $\h _{\mu \nu}$ are
kinematic generators, consequently we concentrate on $\h _\bot$ when
consistency conditions are calculated.

\subsection{Primary if-constraints and super-Hamiltonian}
\label{pish}
\noindent
It is necessary to understand the relation between the canonical momenta and
the velocities before one can make $\h _\bot$ more apparent.
The torsion momenta $\pi _{\ol \nu \mu}$ can be decomposed
into four algebraically irreducible parts:
\begin{eqnarray}
   \pi_{\ol\mu\nu}&=&-n_\nu\pi_{\ol\mu\bot}+\pi_{\ol{\mu\nu}}\nonumber\\
    &=&-n_\nu\pi_{\ol\mu\bot}+{\anti\pi}_{\ol{\mu\nu}}+
       {\symm\pi}_{\ol{\mu\nu}} + {1\over3}\eta_{\ol{\mu\nu}}\pi,
\end{eqnarray}
where the notations refer respectively to the antisymmetric,
symmetric-traceless and the trace parts.
Manipulating the definition of the torsion momenta (\ref{tm}), the following
relations between the different parts of the canonical momenta and the
corresponding parts of the velocities $T_{\bot \ol \mu \nu}$ are found:
\begin{eqnarray}
   \phi _{\ol \mu \bot }&\equiv &{\pi _{\ol \mu \bot }\over J}+
    {1\over 3l^2}(a_1-a_2){\vect T}_{\ol \mu}={1\over 3l^2}(2a_1+a_2)
    T_{\bot \ol \mu \bot },\label{mou4}\\
   {\anti \phi }_{\ol {\mu \nu }}&\equiv &{{\anti \pi }
    _{\ol {\mu \nu }}\over J}+{1\over 3l^2}(a_1-a_3)T_{\ol {\mu \nu }\bot }
    ={1\over 3l^2}(a_1+2a_3)T_{\bot [\ol {\mu \nu}]},\\
   {\symm \phi }_{\ol {\mu \nu }}&\equiv&{{\symm\pi}_{\ol{\mu\nu}}\over J}
    ={a_1\over l^2}T_{\bot \lan \ol {\mu \nu}\ran},\\
   \phi &\equiv &{\pi \over J}={a_2\over l^2}T_{\bot \ol \s}{}^{\ol\s},
\end{eqnarray}
where ${\vect T}_{\ol \mu}\equiv T_{\ol{\mu \nu}}{}^{\ol \nu}$,
and a tensor with two indices contained in the bracket $\lan \, \ran$
denotes that the tensor is symmetric-traceless with respect to the two indices.
If the parameters take on any of the critical values: $2a_1+a_2=0$, $a_1+2a_3=0$,
$a_1=0$, or $a_2=0$, the Lagrangian becomes singular with respect
to some velocities ${\dot\th}_a{}^\mu$ (or equivalently, with respect to
$T_{\bot\ol\nu}{}^\mu$), thus one obtains the following PIC:
$\phi _{\ol \mu \bot}\approx 0$, ${\anti \phi}_{\ol{\mu \nu}}\approx 0$,
${\symm \phi}_{\ol{\mu \nu}}\approx 0$, and/or $\phi \approx 0$, respectively.

The curvature momenta $\pi _{\ol \s \mu \nu}$ can be decomposed into
six irreducible parts:
\begin{equation}
   \pi_{\ol\s\mu\nu}=\pi_{\ol{\s\mu\nu}}+2\pi_{\ol\s\bot[\ol\mu}n_{\nu]},
\end{equation}
and
\begin{eqnarray}
   \pi_{\ol{\mu\nu}\bot}&=&{\anti\pi}_{\ol{\mu\nu}\bot}
    +{\symm\pi}_{\ol{\mu\nu}\bot}+{1\over3}\eta_{\ol{\mu\nu}}\pi_\bot,\\
   \pi_{\ol{\s \mu \nu}}&=&-{1\over 6}\e _{\s \mu \nu\bot}{\ps \pi}
   +{\vect \pi}_{[\ol \mu}\eta _{\ol \nu ]\ol \s}+{4\over 3}
   {\tens \pi}_{\ol \s [\ol{\mu \nu}]}.
\end{eqnarray}
Identifying the irreducible parts of the curvature momenta (\ref{tm}), one finds
\begin{eqnarray}
   \ps\phi &\equiv &{\ps \pi \over J}-{1\over \k}(b_2-b_3){\ps R}_{\circ \bot}
    ={1\over \k}(b_2+b_3){\ps R}_{\bot \circ},\label{323a}\\
   {\vect \phi}_{\ol \mu}&\equiv &{{\vect \pi}_{\ol \mu}\over J}
    +{1\over \k}(b_4-b_5)R_{\ol \mu \bot}=-{1\over \k}(b_4+b_5)R_{\bot \ol \mu},\\
   {\tens \phi}_{\ol{\s \mu \nu}}&\equiv &{{\tens \pi}
    _{\ol{\s \mu \nu}}\over J}+{1\over \k}(b_1-b_2){\tens R}_{\ol{\mu \nu \s}\bot}
    ={1\over \k}(b_1+b_2){\tens R}_{\bot \ol{\s \mu \nu}},\\
   \phi _\bot &\equiv &{\pi _\bot \over J}-{3a_0\over l^2}
    -{1\over 2\k}(b_4-b_6)\ul R={1\over \k}(b_4+b_6)R_{\bot \bot},\label{scm} \\
   {\anti \phi}_{\ol{\mu \nu}\bot}&\equiv &{{\anti \pi}_{\ol{\mu \nu}\bot}\over J}
    -{1\over \k}(b_2-b_5){\ul R}_{[\ol{\mu \nu}]}={1\over \k}(b_2+b_5)
    R_{\bot [\ol{\mu \nu}]\bot},\\
   {\symm \phi}_{\ol{\mu \nu}\bot}&\equiv &{{\symm \pi}_{\ol{\mu \nu}\bot}\over J}
    -{1\over \k}(b_1-b_4){\ul R}_{\lan \ol{\mu \nu}\ran}
    ={1\over \k }(b_1+b_4)R_{\bot \lan \ol{\mu \nu}\ran \bot},\label{323f}
\end{eqnarray}
where ${\ps R}_{\circ \bot}:=\e ^{\mu \nu \s \bot}R_{\ol {\mu \nu
\s}\bot}$,
${\ps R}_{\bot \circ}:=\e ^{\mu \nu \s \bot}R_{\bot \ol {\mu \nu
\s}}$,  $\ul R_{\ol{\mu \nu}}:=R_{\ol{\s \mu \nu}}{}^{\ol \s}$, and
$\ul R:=\ul R_{\ol \mu}{}^{\ol \mu}$.
By a similar argument as used above, for various degenerate parameter combinations
one can obtain any of the six expressions of (\ref{323a}-\ref{323f}) as PIC's.
The relations between the critical parameter combinations and the constraints
are summarized in Table \ref{cpc}.

\begin{table}[htbp]
\tcaption{Primary if-constraints, critical parameter values and masses}
\label{cpc}
\centerline{\footnotesize\smalllineskip
\begin{tabular}{ccllll}\\
\hline\\
 &\multicolumn{2}{l}{Kinetic Parameter}&&&Mass Parameter\\
 \raisebox{1.2ex}[0pt]{$J^p$}&\multicolumn{2}{l}{Combinations}&
 \multicolumn{2}{l}{\raisebox{1.2ex}[0pt]{Constraints}}&Combinations\\
&&&&&\\
 \hline\\
 &(a)&$a_2$&$\phi$,&$\chi$&\\
 \raisebox{1.2ex}[0pt]{$0^+$}&(b)&$b_4+b_6$&$\phi_\bot$,&$\chi_\bot$&
 \raisebox{1.2ex}[0pt]{$a_0$, $2a_0+a_2$}\\
 &(a)&$a_1+2a_3$&${\anti\phi}_{\ol{\mu\nu}}$,&${\anti\chi}_{\ol{\mu\nu}}$&\\
 \raisebox{1.2ex}[0pt]{$1^+$}&(b)&$b_2+b_5$
 &${\anti\phi}_{\ol{\mu\nu}\bot}$,&${\anti\chi}_{\ol{\mu\nu}\bot}$&
 \raisebox{1.2ex}[0pt]{$a_1-a_0$, $\disp{{a_0\over2}+a_3}$}\\
 &(a)&$a_1$&${\symm\phi}_{\ol{\mu\nu}}$,&${\symm\chi}_{\ol{\mu\nu}}$&\\
 \raisebox{1.2ex}[0pt]{$2^+$}&(b)&$b_1+b_4$
 &${\symm\phi}_{\ol{\mu\nu}\bot}$,&${\symm\chi}_{\ol{\mu\nu}\bot}$&
 \raisebox{1.2ex}[0pt]{$a_0$, $a_1-a_0$}\\
 &(a)&$2a_1+a_2$&$\phi_{\ol\mu\bot}$,&$\chi_{\ol\mu\bot}$&\\
 \raisebox{1.2ex}[0pt]{$1^-$}&(b)&$b_4+b_5$
 &${\vect\phi}_{\ol\mu}$,&${\vect\chi}_{\ol\mu}$&
 \raisebox{1.2ex}[0pt]{$a_1-a_0$, $2a_0+a_2$}\\
 $0^-$&&$b_2+b_3$&$\ps\phi$,&$\ps\chi$&$\disp{{a_0\over 2}+a_3}$\\
 $2^-$&&$b_1+b_2$&${\tens\phi}_{\ol{\s\mu\nu}}$,
 &${\tens\chi}_{\ol{\s\mu\nu}}$&$a_1-a_2$\\
&&&&&\\
\hline\\
\end{tabular}}
\end{table}

Blagojevi\'c and Nikoli\'c found that, for the generic PGT, there are only
primary and secondary if-constraints (no tertiary constraints).
The Poisson bracket (PB) between either
(i) appropriately paired primaries,
or (ii) a primary and the secondary  generated by its preservation,
was found to be generally non-vanishing, hence they form a
second-class pair.
More specifically, the value of these PB's are just the
(generally non-vanishing) constant ``mass'' parameter combinations of
Table \ref{cpc} plus some field-dependent terms of nonlinear origin.
In Sec.~5 
we will discuss the effects on
the constraint classification due to these nonlinear terms.

In order to treat all such possibilities in a concise way, the singular function
\begin{equation}\label{mou6}
   {\la (x)\over x}\equiv \left\{ \begin{array}{ll}
    \disp{1\over x}, & x\ne 0, \\
   0, &x=0,\end{array} \right.
\end{equation}
was introduced.
The PIC's in the total Hamiltonian density (\ref{thf}) can be given in the form
\begin{equation}
   u^A\phi _A=(u\cdot \phi)^T+(u\cdot \phi)^R,
\end{equation}
where
\begin{eqnarray}
   (u\cdot \phi)^T&\equiv &[1-\la (2a_1+a_2)]u^{\ol \mu \bot}
    \phi _{\ol \mu \bot}+[1-\la (a_1)]{\symm u}{}^{\ol{\mu \nu}}
    {\symm \phi}_{\ol{\mu \nu}}\nonumber \\
   &&+[1-\la (a_1+2a_3)]{\anti u}{}^{\ol{\mu \nu}}
    {\anti \phi}_{\ol{\mu \nu}}+{1\over 3}[1-\la (a_2)]u\phi\label{mou5}
\end{eqnarray}
and
\begin{eqnarray}
   (u\cdot \phi)^R&\equiv &{1\over 6}[1-\la (b_2+b_3)]\ps u\ps \phi
    +{4\over 3}[1-\la (b_1+b_2)]{\tens u}{}^{\ol {\s \mu \nu}}
    {\tens \phi}_{\ol{\s \mu \nu}}\nonumber \\
   &&+[1-\la (b_4+b_5)]{\vect u}{}^{\ol \mu}{\vect \phi}_{\ol \mu}
    +2[1-\la (b_2+b_5)]{\anti u}{}^{\ol{\mu \nu}\bot}
    {\anti \phi}_{\ol{\mu \nu}\bot}\nonumber \\
   &&+2[1-\la (b_1+b_4)]{\symm u}{}^{\ol{\mu \nu}\bot}{\symm \phi}
    _{\ol{\mu \nu}\bot}+{2\over 3}[1-\la (b_4+b_6)]u^\bot\phi _\bot.
\end{eqnarray}
The super-Hamiltonian $\h _\bot$ in $\h_\rmc$ (\ref{chd}) then
turns out to be of the form
\begin{equation}\label{superh}
   \h _\bot=\h ^T_\bot +\h ^R_\bot,
\end{equation}
with
\begin{eqnarray}
   \h ^T_\bot&=&J{l^2\over 2}\left[ {3\la (2a_1+a_2)\over 2a_1+a_2}
    \phi _{\ol \mu \bot}\phi ^{\ol \mu \bot}
    +{3\la (a_1+2a_3)\over a_1+2a_3}{\anti \phi}_{\ol{\mu \nu}}
    {\anti \phi}{}^{\ol{\mu \nu}}\right.\nonumber \\
   &&\left.\quad +{\la (a_1)\over a_1}{\symm \phi}_{\ol{\mu \nu}}
    {\symm \phi}{}^{\ol{\mu \nu}}+{\la (a_2)\over 3a_2}\phi ^2\right] +
    J{\ul\lg}^T-n^\mu\na _a\pi ^a{}_\mu,\label{mou7}\\
   \h ^R_\bot&=&J\k\left[ {\la (b_2+b_3)\over 24(b_2+b_3)}{\ps \phi}^2
    +{\la (b_4+b_5)\over 4(b_4+b_5)}{\vect \phi}_{\ol \mu}
    {\vect \phi}{}^{\ol \mu}\right.\nonumber \\
   &&\quad+{\la (b_1+b_2)\over 3(b_1+b_2)}{\tens \phi}_{\ol{\s \mu \nu}}
    {\tens \phi}{}^{\ol{\s \mu \nu}}+{\la (b_2+b_5)\over 2(b_2+b_5)}
    {\anti \phi}_{\ol{\mu \nu}\bot}{\anti \phi}{}^{\ol{\mu \nu}\bot}
    \nonumber \\
   &&\left.\quad+{\la (b_1+b_4)\over 2(b_1+b_4)}
    {\symm \phi}_{\ol{\mu \nu}\bot}{\symm \phi}{}^{\ol{\mu \nu}\bot}
    +{\la(b_4+b_6)\over6(b_4+b_6)}\phi_\bot\phi^\bot\right]+J{\ul\lg}^R,
\end{eqnarray}
where
\begin{eqnarray}
   {\ul\lg}^T&=&{1\over 12l^2}\left[ (2a_1+a_3)T_{\ol{\nu \s}\mu}
    T^{\ol{\nu \s}\mu}\right.\nonumber \\
   &&\left.\qquad\, +2(a_1-a_3)T_{\ol{\nu \s \mu}}T^{\ol{\mu \s \nu}}
    -2(a_1-a_2){\vect T}_{\ol \mu}{\vect T}{}^{\ol \mu}\right] ,\\
   {\ul\lg}^R&=&-c_0R_{\ol{\tau \s}\mu \nu}R^{\ol{\tau \s}\mu \nu}
    -c_1R_{\ol{\tau \s \mu}\nu}R^{\ol{\tau \mu \s}\nu}\nonumber \\
   &&-c_2R_{\ol{\tau \s \mu \nu}}R^{\ol{\mu \nu \tau \s}}
    -c_3({\ul R}_{\ol{\mu \nu}}{\ul R}^{\ol{\mu \nu}}+R_{\ol \mu \bot}
    R^{\ol \mu \bot})\nonumber \\
   &&-c_4{\ul R}_{\ol{\mu \nu}}{\ul R}^{\ol{\nu \mu}}
    -c_5{\ul R}^2-{a_0\over 2l^2}{\ul R}+\La,
\end{eqnarray}
here $\La$ is the cosmological constant.
The relations between the constants $c_i$'s and the $b_i$'s are given by
\begin{eqnarray}
   c_0&=&-{1\over 24\k}(2b_1+3b_2+b_3),\nonumber\\
   c_1&=&-{1\over 6\k}(b_1-b_3),\nonumber\\
   c_2&=&-{1\over 24\k}(2b_1-3b_2+b_3),\nonumber\\
   c_3&=&{1\over 4\k}(b_1+b_2-b_4-b_5),\nonumber\\
   c_4&=&{1\over 4\k}(b_1-b_2-b_4+b_5),\nonumber\\
   c_5&=&-{1\over 24\k}(2b_1-3b_4+b_6).
\end{eqnarray}
We now can use the formulations above to construct the total Hamiltonian
under different critical parameter values.
For example, if $2a_1+a_2$ vanishes, this results in the relation in
Eq.~(\ref{mou4}) becomes the constraint
$\phi_{\ol\mu\bot}\equiv\disp{\pi_{\ol\mu\bot}\over J}+\disp{a_1\over l^2}
{\vect T}_{\ol\mu}\approx0$.
Therefore $[1-\la(2a_1+a_2)]u^{\ol\mu\bot}\phi_{\ol\mu\bot}=
u^{\ol\mu\bot}\phi_{\ol\mu\bot}$  in the torsion
if-constraint part (\ref{mou5});
at the same time,
$\disp{3\la(2a_1+a_2)\over2a_1+a_2}\phi_{\ol\mu\bot}\phi^{\ol\mu\bot}=0$ in
$\h^T_\bot$ (\ref{mou7}), by referring to the singular function
(\ref{mou6}).
On the other hand, if $2a_1+a_2\ne0$, the relation (\ref{mou4}) stands and
$\disp{3\la(2a_1+a_2)\over2a_1+a_2}\phi_{\ol\mu\bot}\phi^{\ol\mu\bot}=
\disp{3\over2a_1+a_2}\phi_{\ol\mu\bot}\phi^{\ol\mu\bot}$ and
$[1-\la(2a_1+a_2)]u^{\ol\mu\bot}\phi_{\ol\mu\bot}=0$.
This process can be performed term by term with different critical
parameter combinations.
It is very beneficial for the analysis of the PGT with general or specific
values of the parameters.

\section{The Linearized Theories of PGT}
\label{rab2}
\noindent
It is difficult to tell whether the PGT is viable under physically
reasonable circumstances because of its complications
(Indeed deep analysis has still left the issue obscure).
Accordingly, the weak field approximation approach was
the first and preliminary method applied to the PGT to study its properties.
Linearization gives us insight into the viability of the propagating modes
in the PGT.
In the linearized theory we usually impose the physical requirements that the
propagating modes carry positive energy and do not propagate faster than light.
Theories with higher spin interactions are notorious for having such
problems\cite{Vel269}.
It was noted that interactions involving spins $\ge1$ were
especially vulnerable to having their characteristics outside the metric light
cone, thereby permitting acausal modes of signal propagation\cite{Vel269}.
But linearization gets rid of nonlinear terms.
One can tentatively forget the complex and focus on the structure of the
theory.
It is efficient for quickly eliminating the obviously unfit theories.

In the linearized PGT, in addition to the graviton there are
possible propagating modes with spin$^{\hbox{parity}}= J^p=2^+$, $2^-$,
$1^+$, $1^-$, $0^+$, $0^-$.
The requirements for the possible propagating ``massive'' modes are
essentially $v<c$ (no tachyons) and $E\ge0$ (no ghosts).
Sezgin and van Nieuwenhuizen\cite{SENP80,SezE81} studied the linearized PGT
in the early 80's.
They applied the so-called ``spin-projection operator formalism''\cite{NieP73}
to the PGT and discovered that at least three of
the massive modes have to be eliminated.
By their analysis there are twelve such solutions.
Their results are listed in Table \ref{ti2}.
Hayashi and Shirafuji used the weak field approximation and the Klein-Gordon
equation to investigate the particle spectrum of PGT\cite{HKST80}.
They gave tables of all the maximal dynamic torsion modes.
Their analysis confirmed that these requirements preclude the possibility
of more than 3 dynamic propagating torsion modes.
Consequently, there are many distinct cases with the torsion
modes showing different behavior:  some modes frozen and some dynamic.

\begin{table}[htbp]
\tcaption{Parameter choices for ghost- and tachyon-free gravity Lagrangian
determined by Sezgin and van Nieuwenhuizen.}
\label{ti2}
\centerline{\footnotesize \smalllineskip
\begin{tabular}{clc}\\
\hline\\
  &Parameter choices&Particle content\\
&&\\
\hline\\
  (1)&$a_1+2a_3=b_1+b_2=b_4+b_5=0$&$2^+,0^+,0^-$\\
  (2)&$a_1+2a_3=b_1+b_2=b_1+b_4=0$&$1^-,0^+,0^-$\\
  (3)&$a_1+2a_3=b_1+b_2=b_4+b_6=0$&$2^+,1^-,0^-$\\
  (4)&$2a_1+a_2=b_1+b_2=b_1+b_4=0$&$1^+,0^+,0^-$\\
  (5)&$2a_1+a_2=b_1+b_2=b_4+b_6=0$&$2^+,1^+,0^-$\\
  (6)&$a_1=b_1+b_2=b_4+b_5=0$&$1^+,0^+,0^-$\\
  (7)&$a_1=b_1+b_2=b_4+b_6=0$&$1^+,1^-,0^-$\\
  (8)&$a_1+2a_3=b_2+b_3=b_1+b_4=0$&$2^-,1^-,0^+$\\
  (9)&$2a_1+a_2=b_2+b_3=b_1+b_4=0$&$2^-,1^+,0^+$\\
  (10)&$2a_1+a_2=b_2+b_5=b_1+b_4=0$&$2^-,0^+,0^-$\\
  (11)&$a_1=a_2=b_1+b_2=0$&$1^+,0^-$\\
  (12)&$b_1+b_2=b_4+b_5=b_2+b_5=0$&$0^+,0^-$\\
&&\\
\hline\\
\end{tabular}}
\end{table}

The linearized method was also applied to the cases with ``massless''
modes\cite{SezE81,MNOT81,BRTM85}.
The propagating massless modes can then have gauge freedom.
Kuhfuss and Nitsch\cite{KRNJ86} obtained some very restrictive
PGT parameter restrictions by using the spin-projection operator formalism
and a ``no fourth order pole requirement''.
Their restrictions are
\begin{eqnarray}
&&\La=2a_1+a_2=a_1+2a_2=0,\nonumber\\
&&b_1+b_4=b_4+b_6=b_2+b_5=0,\label{rab1}\\
&&a_1=a_0>0,\quad b_2,b_3\in\mbox{R (arbitrary)}.\nonumber
\end{eqnarray}
This parameter choice basically activates the spin-$0^-$ and spin-$2^-$ modes
(refer to Table \ref{cpc}) and suppresses all the other modes\footnote{Our
understanding of the propagating modes from this parameter choice is different
from Ref.~1, 
further study is needed to clarify this point.}.
These two propagating modes are both massless.

The idea of ``second-class pairs'' of constraints introduced in
Sec.~3 
is useful in analyzing the linearized PGT.
It shows that the Hamiltonian analysis is in complete accord with
the linearized theory analysis of the propagating
modes\cite{SENP80,HKST80,SezE81}.
The cases where one or more of the ``masses'' vanish have also been analyzed
by Blagojevi\'c's group using the Hamiltonian procedure---although only
to linear order\cite{BMVM87}.
The results are in accord with the linearized theory analysis of propagating
massless modes and their associated extra gauge
symmetries\cite{KRNJ86,SezE81,MNOT81,BRTM85}.

It is satisfying that all of the aforementioned results are essentially
consistent with
one another, however, we should beware of the limitations of linearization.
Although we may obtain a relatively quick understanding of the system via
the linearization of the theory, the result from the full theory,
especially a highly nonlinear system, does not necessarily keep the same
traits as its linearization.
This fact has made us suspicious regarding these pioneers' proclamations.
Especially, we view the conclusions of linearized theory for the
massless modes as rather likely requiring qualitative revision from
nonlinear effects and hence we regard them as much less firm than those
for the massive modes.
Consequently, we don't know how much confidence to place in their conclusions.
Our reservations concerning the linearized propagating massless modes in no way
affect the main conclusions from the survey of the linearized PGT.
However, it seems likely that including the nonlinear effect terms would
modify the Poisson brackets of the constraints,
converting many of the vanishing brackets to non-vanishing ones.
In this way a constraint which was first-class, with
vanishing (on shell) brackets with all other constraints, would move to the
second-class indicating that a gauge symmetry of the linearized theory was
broken in the nonlinear theory.
Hence we expect that the present understanding of ``massless'' PGT modes
and their extra gauge symmetries, being based on a linearized approximation
which appears inadequate, will require substantial revisions.

\begin{table}[htbp]
\tcaption{No ghosts and no tachyons conditions for the torsion modes}
\label{ti1}
\centerline{\footnotesize \smalllineskip
\begin{tabular}{ccl}\\ 
\hline\\
  $J^p$&$\qquad$&Energy$>0$ and Mass${}^2>0$\\
&&\\
\hline\\
  $0^+$&$\qquad$&$b_4+b_6>0,\quad a_0a_2(a_2+2a_0)<0$\\
  $0^-$&$\qquad$&$b_2+b_3<0,\quad a_3+a_0/2<0$ \\
  $1^+$&$\qquad$&$b_2+b_5>0,\quad (a_1+2a_3)(a_1-a_0)(a_3+a_0/2)>0$\\
  $1^-$&$\qquad$&$b_4+b_5<0,\quad (2a_1+a_2)(a_1-a_0)(a_2+2a_0)<0$\\
  $2^+$&$\qquad$&$a_0<0,\quad b_1+b_4>0,\quad a_0a_1(a_1-a_0)>0$\\
  $2^-$&$\qquad$&$b_1+b_2<0,\quad a_1-a_0<0$\\
&&\\
\hline\\
\end{tabular}}
\end{table}

Aside from nonlinear effects, the analyses of the linearized theories of
PGT offers the identification of the torsion modes, the critical kinetic
parameter combinations, the mass terms and other useful relations.
The critical kinetic and mass parameter combinations are given in
Table \ref{cpc}.
The necessary signature conditions for no ghosts and no tachyons are
presented in Table \ref{ti1}.

\section{Hamiltonian Analysis of Propagating Torsion Modes}
\label{haptm}
\noindent
The works discussed in the last section  showed that there are
still many viable linearized theories of PGT.
In the other words, the linearized theories with these parameter
choices have good behavior in the weak field approximation. The next step is
to consider if the full nonlinear theories with the same parameter choices
still survive. It is quite important that the whole structure of the
theory stays the same
before and after linearization.
 (Theories that do not have this character are
inherently nonlinear, and are much more difficult to analyze.  We are very
reluctant to seriously entertain such theories unless we have no
alternative.)
In particular we should have no change
in the number and type of constraints, for that is what determines the
number of gauge freedoms and the number of dynamic physical degrees
of freedom.
If the answer is affirmative, the theory can be smoothly transformed from
nonlinearity to linearity and vice versa with its properties unchanged.
If the answer is negative, the properties of the theory would change
fundamentally upon linearization.
In fact
it has been argued that such changes in the properties are  accompanied
by the occurrence of serious physical problems, in particular anomalous
characteristics\cite{HRNZ94,Chen98}.
Hence a sound case of the PGT with ``good'' parameter choices should not
exhibit such a change of properties behavior
when passing to the weak field
approximation. Thus this becomes a criterion
that determines if some specific parameter choice is good enough, or
whether it should be regarded as non-viable.

We are going to investigate the viability of the individual simple propagating
spin modes and two better-behaved cases.
Here ``simple'' indicates that all parameters will be set to vanish except the
key ones.
Such severe conditions could be generalized.
For example, in order to suppress all except the $0^+$ and $0^-$ modes,
according Table \ref{cpc}.
our condition could be relaxed to require only $b_1=-b_2=-b_4=b_5$
instead of our severe ``simple'' choice
$b_1=b_2=b_4=b_5=0$.
There are two reasons not to adopt the ``relaxed'' versions:
(a) it would much complicate the analysis,
(b) the relaxed parameter choices do not essentially enhance the viability
of the theory; instead, they could even ruin a case which is well-behaved
under the simple version.
We can readily see this from the primary Poisson matrices of all the cases.

It is not the purpose of this paper to examine viability in the full
nonlinear domain of all the ``viable'' linearized theories of PGT.
It seems appropriate to first try to illustrate
the influence of nonlinearity in these theories and to investigate the
effects in order to get some understanding.  Nevertheless, we believe
that   the phenomena resulting from nonlinearity
in the cases discussed in this paper are representative for all the other cases.
We would like to emphasize the effects of nonlinearity on the change of the
structure and gauges of the theory, so we will focus on the change
of the type and/or number of the constraints and the consistency of the whole
system caused by nonlinearity.
These observations will be solid enough for an understanding of the
differences before and after linearization.

In our investigations we found that the  ``constraint
bifurcation''  phenomena is
a key evidence which determines if there is an anomalous characteristic.
Whether the Poisson matrix has constant rank or not, independent of the
field values, is the criteria for
knowing whether the phenomenon of constraint bifurcation occurs.
A field dependent rank can happen only if the constraints are nonlinear.
Strange behavior may occur as one approaches a point in the phase
space where the rank of the Poisson matrix changes.
At such points the number and type of constraints (and thus the number of
gauges and physical degrees of freedom) are not constant.
This phenomenon signals the presence of an acausal propagating mode in the
system and is unacceptable theoretically\cite{Chen98}.
Whether the Poisson matrix is singular can be judged by the positivity
of its determinant.
Since the matrix is usually big the concise form of its determinant
is hard to obtain, so we appeal to the aid of the symbolic computing
software {\tt REDUCE}\cite{HeaA85}
for the calculations.
 From the result associated with all the constraints we can decide the
positivity of the determinant and thus the possibility of  constraint
bifurcation.

\subsection{Simple spin-one and spin-two cases}
\noindent
The method of the analysis is simple conceptually:
list all of the constraints and compare the differences of the
 structure before
and after linearization.
The Poisson matrix is formed containing the non-zero Poisson Brackets
(PB), then checked for change under linearization.
Due to the complications, the secondary constraints (SC) will not be
expressed explicitly.
Instead we will reason out the differences based on the result of the SC's
linearization in certain situations.

\subsubsection{The simple spin-$1^+$ case}
\noindent
${\anti\pi}_{\ol{\mu\nu}\bot}$ corresponds to the spin-$1^+$ mode.
The linearized theory indicates that the theory with only a spin-$1^+$
mode can be obtained with  the simple parameter choice:
\begin{equation}\label{sp1p1}
   \begin{array}{l}
      a_0\ne 0,\quad a_3\ne0,\quad b_5\ne 0,\\
      a_1=a_2=0,\\
      b_1=b_2=b_3=b_4=b_6=0.
   \end{array}
\end{equation}
With such a parameter choice there are twenty-one PIC's.
The PIC's are as follows:
\begin{equation}
   \begin{array}{rclrclrcl}
      \phi&\equiv&\disp{\pi\over J}\approx 0,\quad&\disp{\symm\phi}
       _{\ol{\mu\nu}}&\equiv&\disp{{\symm\pi}_{\ol{\mu\nu}}\over J}\approx0,
       \quad&\phi_{\ol\mu\bot}&\equiv&\disp{\pi_{\ol\mu\bot}\over J}\approx0,\\
      \ps\phi&\equiv&\disp{\ps\pi\over J}\approx 0,\quad
       &{\tens\phi}_{\ol{\s\mu\nu}}&\equiv&\disp{{\tens\pi}_{\ol{\s\mu\nu}}
       \over J}\approx0,&&&\\
      \phi_\bot&\equiv&\disp{\pi\over J}-\disp{3a_0\over l^2}\approx0,\quad
       &{\symm\phi}_{\ol{\mu \nu}\bot}&\equiv&\disp{{\symm\pi}_{\ol{\mu\nu}\bot}
       \over J}\approx0.&&&
   \end{array}
\end{equation}
The non-zero PB's are
\begin{eqnarray}
   \{\phi_{\ol\mu\bot},\phi_{\ol\mu\bot}^\prime\}
    &=&2{\de_{xx^\prime}\over J^2}{\anti\pi}_{\ol{\mu\nu}},\label{s1pd}\\
   \{{\symm\phi}_{\ol{\mu\nu}},{\symm\phi}{}_{\ol{\tau\s}}^\prime\}
    &=&2{\de_{xx^\prime}\over J^2}{\anti\pi}_{(\ol\tau(\ol\mu}
    \eta_{\ol\nu)\ol\s)},\\
   \{{\ps\phi},\phi_{\ol\mu\bot}^\prime\}
    &=&2{\de_{xx^\prime}\over J^2}\e_\mu{}^{\nu\s\bot}
    {\anti\pi}_{\ol{\nu\s}\bot},\\
   \{\phi_\bot,\phi^\prime\}
    &=&-{\de_{xx^\prime}\over J}{6a_0\over l^2},\label{s1pa}\\
   \{\phi_\bot,\phi_{\ol\mu\bot}^\prime\}
    &=&-{\de_{xx^\prime}\over J^2}{\vect\pi}_{\ol\mu},\\
   \{{\tens\phi}_{\ol{\s\mu\nu}},\phi_{\ol\tau\bot}^\prime\}
    &=&{\de_{xx^\prime}\over J^2}\tens(\eta_{\ol{\tau\nu}}
    {\anti\pi}_{\ol{\s\mu}\bot})\\
   \{{\tens\phi}_{\ol{\s\mu\nu}},{\symm\phi}{}_{\ol{\rho\tau}}^\prime\}
    &=&{1\over2}{\de_{xx^\prime}\over J^2}\left[{\vect\pi}_{\ol\mu}
    \eta_{\ol\nu\lan\ol\rho}\eta_{\ol\tau\ran\ol\s}
    -\tens({\vect\pi}_{\ol\s}\eta_{\ol\mu\lan\ol\rho}\eta_{\ol\tau\ran\ol\nu})
    \right],\label{s1pc}\\
   \{{\symm\phi}_{\ol{\mu\nu}\bot},\phi_{\ol\s\bot}^\prime\}
    &=&{1\over2}{\de_{xx^\prime}\over J^2}{\vect\pi}_{\lan\ol\mu}
    \eta_{\ol\nu\ran\ol\s},\\
   \{{\symm\phi}_{\ol{\mu\nu}\bot},{\symm\phi}{}_{\ol{\tau\s}}^\prime\}
    &=&{\de_{xx^\prime}\over J}\left[{1\over J}\eta_{(\ol\mu(\ol\tau}
    {\anti\pi}_{\ol\s)\ol\nu)}+{a_0\over l^2}\eta_{\ol\tau\lan\ol\mu}
    \eta_{\ol\nu\ran\ol\s}\right]\label{s1pb}.
\end{eqnarray}
With these PB's we can construct the primary Poisson matrix with the
following form:
        \begin{equation}
   \begin{array}{ccc|ccccccc|}
      \multicolumn{3}{c}{}&1&3&5&1&1&5&\multicolumn{1}{c}{5}\\
      \multicolumn{3}{c}{}&\ob{}&\ob{}&\ob{}&\ob{}&\ob{}&\ob{}&
       \multicolumn{1}{c}{\ob{}}\\
      \multicolumn{3}{c}{}&\phi&\phi_{\ol\mu\bot}&{\symm\phi}_{\ol{\mu\nu}}&
       \phi_\bot&\ps\phi&{\symm\phi}_{\ol{\mu\nu}\bot}&
       \multicolumn{1}{c}{{\tens\phi}_{\ol{\s\mu\nu}}}\\
      \multicolumn{10}{c}{}\\
      1&\{&\phi&\bc&\bc&\bc&\eta&\bc&\bc&\bc\\
      3&\{&\phi_{\ol\mu\bot}&\bc&\pi&\bc&\pi&\pi&\pi&\pi\\
      5&\{&{\symm\phi}_{\ol{\mu\nu}}&\bc&\bc&\pi&\bc&\bc&{\eta+}\pi&\pi\\
      1&\{&\phi_\bot&\eta&\pi&\bc&&&&\\
      1&\{&\ps\phi&\bc&\pi&\bc&&&&\\
      5&\{&{\symm\phi}_{\ol{\mu\nu}\bot}&\bc&\pi&{\eta+}\pi&
       \multicolumn{4}{c|}{\raisebox{2.2ex}[5pt]{$\bc$}}\\
      5&\{&{\tens\phi}_{\ol{\s\mu\nu}}&\bc&\pi&\pi&&&&
   \end{array}
\label{ts1d}
\end{equation}
where $\eta$'s represent the constant terms and $\pi$'s represent the
terms involving the variables. The details of these terms can be found
in Eq.~(\ref{s1pd})-(\ref{s1pb}).
The super-Hamiltonian density is deduced by applying the parameter choice
(\ref{sp1p1}) to (\ref{superh}).
It is
\begin{equation}
   \h_\bot^{1^+}=\h_\bot^{R1^+}+\h_\bot^{T1^+},
\end{equation}
where
\begin{eqnarray}
   \h_\bot^{R1^+}
    &=&{1\over2}{\vect\pi}_{\ol\mu}\left[{\k\over2b_5}{{\vect\pi}
    {}_{\ol\mu}\over J}+R^{\ol\mu\bot}\right]
    -{\anti\pi}_{\ol{\mu\nu}\bot}\left[{\k\over2b_5}
    {{\anti\pi}{}^{\ol{\mu\nu}}{}_\bot\over J}+{\ul R}{}^{[\ol{\mu\nu}]}
    \right]-{a_0\over l^2}J{\ul R},\label{s1psH}\\
   \h_\bot^{T1^+}
    &=&{1\over2}{\anti\pi}_{\ol{\mu\nu}}\left[{3l^2\over2a_3}
    {{\anti\pi}{}^{\ol{\mu\nu}}\over J}+T^{\ol{\mu\nu}\bot}\right]
    +{a_3\over24l^2}J{\ps T}^2-n^\mu\na_a\pi^a{}_\mu.
\end{eqnarray}
Within the phase surface of primary constraints, the super-momenta
(\ref{superm}) and the Lorentz rotation parts (\ref{lorentzp}) lead to
the following constraints:
\begin{eqnarray}
   \h_{\ol\mu}^{1^+}
    &\lr&\eta_{\ol{\mu\nu}}\na_{\ol\s}{{\anti\pi}{}^{\ol{\nu\s}}\over J}
    +R_{\ol{\mu\nu\s}\bot}{{\anti\pi}{}^{\ol{\nu\s}\bot}\over J}
    -{1\over2J}{\ul R}_{\ol{\mu\nu}}{\vect\pi}{}^{\ol\nu}
    -{a_0\over l^2}R_{\ol\mu\bot} \nonumber\\
   &&+{3\over2J}{\anti\pi}_{\ol{\mu\nu}}{\vect T}{}^{\ol\nu}-{2\over3J}
    {\tens T}_{\ol{\nu\s\mu}}{\anti\pi}{}^{\ol{\nu\s}}-{1\over6J}
    \e_{\mu\nu\s\bot}{\ps T}{\anti\pi}{}^{\ol{\nu\s}}\approx0,\\
   \h_{\ol\mu\bot}^{1^+}
    &\lr&{{\anti\pi}_{\ol{\mu\nu}\bot}\over J}{\vect T}{}^{\ol\nu}
    -{a_0\over l^2}{\vect T}_{\ol\mu}+{1\over J}{\vect\pi}_{[\ol\mu}
    \na_{\ol\nu]}n^\nu-\de_{\ol\mu}{}^\s\na_{\ol\nu}{{\anti\pi}{}^{\ol\nu}
    {}_{\ol\s\bot}\over J}\approx0,\\
   \h_{\ol{\mu\nu}}^{1^+}
    &\lr&{{\anti\pi}_{\ol{\mu\nu}}\over J}+{a_0\over2l^2}T_{\ol{\mu\nu}\bot}
    +{1\over2J}{\vect\pi}_{[\ol\mu}{\vect T}_{\ol\nu]}\nonumber\\
   &&\qquad\qquad\qquad +{1\over2}\de^\s{}_{[\ol\mu}\na_{\ol\nu]}
    {{\vect\pi}_{\ol\s}\over J}-{1\over J}{\anti\pi}{}^{\ol\s}{}_{[\ol\mu|\bot}
    \na_{\ol\s}n_{|\nu]}\approx0.
\end{eqnarray}

It is essential to compare the nonlinear spin-$1^+$ case with its linearized
version.
At first let us describe briefly the analysis of the linear spin-$1^+$ case.
After linearization, the remaining non-zero PB's turn out to be
Eqs.~(\ref{s1pa}), (\ref{s1pb}) in which only the zero-order (constant) terms
exist, i.e.,
\begin{eqnarray}
   \{\phi_\bot^\fl,\phi^{\fl\prime}\}
    &\approx&\de_{xx^\prime}{6a_0\over l^2},\label{oplpb1}\\
   \{{\symm\phi}{}_{\ol{\mu\nu}\bot}^\fl,
    {\symm\phi}{}_{\ol{\tau\s}}^{\fl\prime}\}
    &\approx&-\de_{xx^\prime}{a_0\over l^2}\eta_{\ol\tau\lan\ol\mu}^\fl
    \eta_{\ol\nu\ran\ol\s}^\fl,\label{oplpb2}
\end{eqnarray}
where the superscript $\fl$ indicates linearization.
With the linearized PB's we can construct the constant primary
Poisson matrix:
\begin{equation}
   \begin{array}{ccc|ccccccc|}
      \multicolumn{3}{c}{}&1&3&5&1&1&5&\multicolumn{1}{c}{5}\\
      \multicolumn{3}{c}{}&\ob{}&\ob{}&\ob{}&\ob{}&\ob{}&
       \ob{}&\multicolumn{1}{c}{\ob{}}\\
      \multicolumn{3}{c}{}&\phi^\fl&\phi_{\ol\mu\bot}^\fl&
       {\symm\phi}{}_{\ol{\mu\nu}}^\fl&
       \phi_\bot^\fl&\ps\phi^\fl&{\symm\phi}{}_{\ol{\mu\nu}\bot}^\fl&
       \multicolumn{1}{c}{{\tens\phi}_{\ol{\s\mu\nu}}^\fl}\\
      \multicolumn{10}{c}{}\\
      1&\{&\phi^\fl&&&&\eta&\bc&\bc&\bc\\
      3&\{&\phi_{\ol\mu\bot}^\fl&&\bc&&\bc&\bc&\bc&\bc \\
      5&\{&{\symm\phi}{}_{\ol{\mu\nu}}^\fl&&&&\bc&\bc&\eta&\bc\\
      1&\{&\phi_\bot^\fl&-\eta&\bc&\bc&&&&\\
      1&\{&\ps\phi^\fl&\bc&\bc&\bc&&&&\\
      5&\{&{\symm\phi}{}_{\ol{\mu\nu}\bot}^\fl&\bc&\bc&-\eta&
       \multicolumn{4}{c|}{\raisebox{2.2ex}[5pt]{$\bc$}}\\
      5&\{&{\tens\phi}_{\ol{\s\mu\nu}}^\fl&\bc&\bc&\bc&&&&
   \end{array}
\label{ts1dL}
\end{equation}
where $\eta$ represents the constants in Eq.~(\ref{oplpb1}) and
(\ref{oplpb2}).
The consistency conditions guarantee that the linearized Lagrange multipliers,
i.e., $u^\fl$, $u^{\fl\bot}$, ${\symm u}{}^{\fl\ol{\mu\nu}}$,
and ${\symm u}{}^{\fl\ol{\mu\nu}\bot}$, are obtained.
The nine linearized SC, $\chi_{\ol\mu\bot}^\fl$, $\ps\chi^\fl$ and
${\tens\chi}{}_{\ol{\s\mu\nu}}^\fl$, also show up when the consistency
conditions of the linearized constraints $\phi_{\ol\mu\bot}^\fl$,
${\ps\phi}{}^\fl$ and ${\tens\phi}{}_{\ol{\s\mu\nu}}^\fl$ are required.
Furthermore, the linearized SC $\chi_{\ol\mu\bot}^\fl$, ${\ps\chi}^\fl$ and
${\tens\chi}{}_{\ol{\s\mu\nu}}^\fl$ will become the counterparts to the
second-class pairs of $\phi_{\ol\mu\bot}^\fl$, ${\ps\phi}{}^\fl$ and
${\tens\phi}{}_{\ol {\s\mu\nu}}^\fl$ respectively according to
Ref.~23. 
These second-class pairs determine the multipliers $u^{\fl\ol\mu\bot}$,
${\ps u}{}^\fl$ and ${\symm u}{}^{\fl\ol{\mu\nu}}$ and the algorithm of
the linearized case is terminated.
In the PGT, there are forty dynamic variables coming from the sixteen
tetrad components and the twenty-four connection components.
The number of total variables counts eighty because the same number of
canonical momenta accompany the variables.
Nonetheless, constraints eliminate many unphysical variables.
It is known that the super-Hamiltonian $\h_\bot$, super-momenta $\h _a$,
and the Lorentz generators $\h_{\mu\nu}$ are ten first-class constraints.
There are also ten sure first-class primary constraints.
The total twenty first-class constraints (because of their gauge nature)
offset forty variables.
The degrees of freedom of the linearized case can be counted as
${1\over2}(80[\mbox{total}]-40[\mbox{gauge}]-21[\mbox{PIC}]-9[\mbox{SC}])
=5=3[1^+]+2[\mbox{GR}]$, a massive spin-$1^+$ field in addition of the
usual graviton.
For the sake of ``ghost-free'' the positivity of the kinetic energy density
demands $a_3<0$ and $b_5/\k>0$.

Notwithstanding the good behavior of the linearized case,
the nonlinearity of the full spin-$1^+$ theory ruins its whole legitimacy.
Generically, the Lagrange multipliers $u$, $u^\bot$,
${\symm u}{}^{\ol{\mu\nu}}$, ${\symm u}{}^{\ol{\mu\nu}\bot}$
can not be determined alone because the nonlinear terms are engaged
(see Eq.~(\ref{ts1d})).
Here we assume tentatively that there is no inconsistency appearing
in the system during the process.
The consistency conditions of the connection PIC's, $\phi_\bot$, $\ps\phi$,
${\symm\phi}_{\ol{\mu\nu}\bot}$ and ${\tens\phi}_{\ol{\s\mu\nu}}$,
will determine the multipliers of the tetrad PIC's, $u$, $u^{\ol\mu\bot}$
and ${\symm u}{}^{\ol {\mu\nu}}$, and give three SC's, $\chi_{[3]}$.
Principally, the consistency conditions of the tetrad PIC's and the SC's
$\chi_{[3]}$ will determine the multipliers, $u^\bot$, $\ps u$,
${\symm u}{}^{\ol{\mu \nu}\bot}$ and ${\tens u}{}^{\ol{\s\mu\nu}}$
 according to the experience of the linearized case and terminate the process.
The degrees of freedom of the full nonlinear case count as
${1\over2}(80-40-21[\mbox{PIC}]-3[\mbox{SC}])=8=2+3[1^+]+3[1^-]$.
This shows that the massive spin-$1^-$ field is excited as well as
the massive spin-$1^+$ field since the other modes are already suppressed.
The result is unacceptable from the point of view of the positivity of the
kinetic energy density in the Hamiltonian density (refer to Eq.~(\ref{s1psH})).

One can learn the major difference just from the primary Poisson matrix.
The rank of the linearized primary Poisson matrix (\ref{ts1dL}) in this case
is twelve, a constant, whereas the rank of the full nonlinear matrix
(\ref{ts1d}) varies.
Its rank is eighteen generically but it can be less than eighteen depending
on some specific numerical values of the variables which could degenerate
the system.
We verified this point by using the symbolic
computation software {\tt REDUCE} to calculate the determinant
of the matrix (\ref{ts1d}).
The result shows that the determinant is not positive definite,
which indicates that whether the system is singular depends on the
numerical values of the variables.
This situation shows the phenomenon of constraint bifurcation.
It could even happen as the system dynamically evolves. Consider
all of the possible
initial values that give generic rank to the Poisson matrix.  We see no
 mechanism that would guarantee that every one of these cases could not
evolve to field values where the rank changes.  Because of these features
we regard this parameter choice as physically unacceptable.

\subsubsection{The simple spin-$1^-$ case}
\noindent
We next examine the simple spin-$1^-$ case and find that it has similar
problems as in the simple spin-$1^+$ case.
${\vect\pi}_{\ol\mu}$ corresponds to the spin-$1^-$ mode.
The linearized theory indicates that the theory with only the spin-$1^-$ mode
is obtained if only $a_0$, $a_2$ and $b_5$ do not vanish.
Under this parameter choice ${\anti\phi}_{\ol{\mu\nu}}$ instead of
$\phi$ and $\phi_{\ol\mu\bot}$ become PIC's and the other PIC's remain
just as for the spin-$1^+$ case.
The non-zero PB's are
\begin{eqnarray}
   \{\ps\phi,{\anti\phi}{}_{\ol{\mu\nu}}^\prime\}
    &=&{\de_{xx^\prime}\over J^2}\e_{\mu\nu}{}^{\s\bot}{\vect\pi}_{\ol\s},\\
   \{\phi_\bot,{\anti\phi}{}_{\ol{\mu\nu}}^\prime\}
    &=&{\de_{xx^\prime}\over J^2}{\anti\pi}_{\ol{\mu\nu}\bot},\\
   \{\tens\phi_{\ol{\s\mu\nu}},{\anti\phi}{}_{\ol{\tau\rho}}^\prime\}
    &=&-{\de_{xx^\prime}\over2J^2}\tens(\eta_{\ol\mu[\ol\tau}
    \eta_{\ol\rho]\ol\nu}{\vect\pi}_{\ol\s}),\\
   \{\symm\phi_{\ol{\mu\nu}\bot},{\anti\phi}{}_{\ol{\tau\s}}^\prime\}
    &=&{\de_{xx^\prime}\over J^2}{\anti\pi}_{[\ol\tau\lan\ol\mu|\bot}
    \eta_{|\ol\nu\ran\ol\s]},\\
   \{\tens\phi_{\ol{\s\mu\nu}},{\symm\phi}{}_{\ol{\rho\tau}}^\prime\}
    &=&\mbox{rhs of Eq.~(\ref{s1pc})},\\
   \{\symm\phi_{\ol{\mu\nu}\bot},{\symm\phi}{}_{\ol{\tau\s}}^\prime\}
    &=&\mbox{rhs of Eq.~(\ref{s1pb})}.
\end{eqnarray}
By utilizing the PB's above the primary Poisson matrix has the form
\begin{equation}
   \begin{array}{ccc|cccccccc|}
      \multicolumn{3}{c}{}&1&3&3&5&1&1&5&\multicolumn{1}{c}{5}\\
      \multicolumn{3}{c}{}&\ob{}&\ob{}&\ob{}&\ob{}&\ob{}&\ob{}&
       \ob{}&\multicolumn{1}{c}{\ob{}}\\
      \multicolumn{3}{c}{}&\phi&\phi_{\ol\mu\bot}&{\anti\phi}_{\ol{\mu\nu}}&
      {\symm\phi}_{\ol{\mu\nu}}&\phi_\bot&\ps\phi&{\symm\phi}_{\ol{\mu\nu}\bot}&
       \multicolumn{1}{c}{{\tens\phi}_{\ol{\s\mu\nu}}}\\
      \multicolumn{10}{c}{}\\
      1&\{&\phi&&&&&\eta&\bc&\bc&\bc\\
      3&\{&\phi_{\ol\mu\bot}&&&&&\pi&\pi&\pi&\pi\\
      3&\{&{\anti\phi}_{\ol{\mu\nu}}&
       \multicolumn{4}{c}{\raisebox{2.2ex}[5pt]{$\bc$}}&\pi&\pi&\pi&\pi\\
      5&\{&{\symm\phi}_{\ol{\mu\nu}}&&&&&\bc&\bc&\eta+\pi&\pi\\
      1&\{&\phi_\bot&\eta&\pi&\pi&\bc&&&&\\
      1&\{&\ps\phi&\bc&\pi&\pi&\bc&&&&\\
      5&\{&{\symm\phi}_{\ol{\mu\nu}\bot}&\bc&\pi&\pi&\eta+\pi&
       \multicolumn{4}{c|}{\raisebox{2.2ex}[5pt]{$\bc$}}\\
      5&\{&{\tens\phi}_{\ol{\s\mu\nu}}&\bc&\pi&\pi&\pi&&&&
   \end{array}
\end{equation}
The super-Hamiltonian density is
\begin{equation}
   \h_\bot^{1^-}=\h_\bot^{R1^+}+\h_\bot^{T1^-},
\end{equation}
where
\begin{equation}
   \h_\bot^{T1^-}=\pi_{\ol\mu\bot}\left[{3l^2\over2a_2}
    {\pi^{\ol\mu\bot}\over J}+{\vect T}{}^{\ol\mu}\right]+{l^2\over6a_2}
    {\pi^2\over J}-n^\mu\na_a\pi^a{}_\mu.
\end{equation}
The constraints derived from the super-momenta and
the Lorentz rotation parts are:
\begin{eqnarray}
   \h_{\ol\mu}^{1^-}
    &\approx&R_{\ol{\mu\nu\s}\bot}{{\anti\pi}{}^{\ol{\nu\s}\bot}\over J}
    -{1\over2J}{\ul R}_{\ol{\mu\nu}}{\vect\pi}{}^{\ol\nu}\nonumber\\
   &&\qquad\qquad\qquad-{a_0\over l^2}R_{\ol\mu\bot}+{\pi^{\ol\nu}{}_\bot
    \over J}\na_{\ol\mu}n_\nu-{1\over3}\na_{\ol\mu}{\pi\over J}\approx0,\\
   \h_{\ol\mu\bot}^{1^+}
    &\approx&{{\anti\pi}_{\ol{\mu\nu}\bot}\over J}{\vect T}{}^{\ol\nu}
    -{\pi_{\ol\mu\bot}\over J}-{a_0\over l^2}{\vect T}_{\ol\mu}\nonumber\\
   &&\qquad\qquad\qquad+{1\over J}{\vect\pi}_{[\ol\mu}\na_{\ol\nu]}n^\nu
    -\de_{\ol\mu}{}^\s\na_{\ol\nu}{{\anti\pi}{}^{\ol\nu}
    {}_{\ol\s\bot}\over J}\approx0,\\
   \h_{\ol{\mu\nu}}^{1^+}
    &\approx&{a_0\over 2l^2}T_{\ol{\mu\nu}\bot}+{1\over2J}
    {\vect\pi}_{[\ol\mu}{\vect T}_{\ol\nu]}\nonumber\\
   &&\qquad\qquad\qquad +{1\over2}\de^\s{}_{[\ol\mu}\na_{\ol\nu]}
    {{\vect\pi}_{\ol\s}\over J}-{1\over J}{\anti\pi}{}^{\ol\s}{}_{[\ol\mu|\bot}
    \na_{\ol\s}n_{|\nu]}\approx0.
\end{eqnarray}
By using same argument as in the simple spin-$1^+$ case, it is obvious that
nonlinearity remains untamed.
Linearly the degrees of freedom are
${1\over2}(80-40-20[\mbox{PIC}]-10[\mbox{SC}])=5=3[1^-]+ 2[\mbox{GR}]$.
The spin-$1^-$ field is obtained as we expected.
But nonlinearly it becomes
${1\over2}(80-40-20[\mbox{PIC}]-4[\mbox{SC}])=8=3[1^-]+3[1^+] +2$.
The appearance of spin-$1^+$ allows the possibility of negative kinetic energy
and hinders the success of the case.
The possible constraint bifurcation is noted and its consequence is
similar to the spin-$1^+$ case discussed above.

In the aforementioned spin-one cases our analysis shows that neither the
simple
spin-$1^+$ mode nor the simple spin-$1^-$ mode can be separated from each
other.
To see this more clearly, we can consider a wilder parameter arrangement:
$a_0$ and $b_5$ non-vanishing only.
This choice has the consequence that all tetrad variables are not dynamical;
they turn out to be the tetrad PIC's:  $\phi$, $\phi_{\ol\mu \bot}$,
${\anti\phi}_{\ol{\mu\nu}}$ and ${\symm\phi}_{\ol{\mu\nu}}$.
The other PIC's remain the same:  $\ps\phi$, $\phi_\bot$,
${\tens\phi}_{\ol{\s\mu\nu}}$ and ${\symm\phi}_{\ol{\mu\nu}\bot}$.

We find that generically the degrees of freedom count as
${1\over2}(80-40-24[\mbox{PIC}])=8=3[1^-]+3[1^+]+2$.
The only difference between this ``simpler'' case and the simple spin-one cases
discussed above
is that the SC's induced from the PIC's in the simple spin-one cases
have essentially transformed into PIC's coming from the parameter choice
in this ``simpler'' case.
However, the spin-one modes still stick to each other to crash
the theory because of the negative kinetic energy.

\subsubsection{The simple spin-$2^-$ case}
\noindent
With the understanding of the simple spin-one cases the spin-$2^+$ mode which
stands with the parameter choice of only $a_0$, $a_1$ and $b_1$ non-vanishing
is not expected to have good physical propagation because it is entangled
in a more complicated nonlinearity.
Instead we examine the simple spin-$2^-$ case.
Although it eventually becomes futile,
it is illuminating for our next case.
According to Table \ref{cpc}  $\tens\pi_{\ol{\s\mu\nu}}$ corresponds to
the spin-$2^-$ mode.
We make the specific parameter choices:
\begin{equation}
   \begin{array}{l}
      a_0\ne 0,\quad b_1\ne 0,\\
      a_1=a_2=a_3=0,\\
      b_2=b_3=b_4=b_5=b_6=0.
   \end{array}
\end{equation}
The PIC's are as follows:
\begin{eqnarray}
   \phi\equiv{\pi\over J}\approx 0,\quad
    &&{\symm\phi}_{\ol{\mu\nu}}\equiv{{\symm\pi}_{\ol{\mu\nu}}\over J}
    \approx 0,\nonumber\\
   \phi_{\ol\mu\bot}\equiv{\pi_{\ol\mu\bot}\over J}\approx0,\quad
    &&{\anti\phi}_{\ol{\mu\nu}}\equiv{{\anti\pi}_{\ol{\mu\nu}}\over J}
    \approx 0,\nonumber\\
   \ps\phi\equiv{\ps\pi\over J}\approx0,\quad
    &&{\vect\phi}_{\ol\mu}\equiv{{\vect\pi}_{\ol\mu}\over J}\approx0,
    \nonumber\\
   \phi_\bot\equiv{\pi\over J}-{3a_0\over l^2}\approx0,\quad
    &&{\anti\phi}_{\ol{\mu\nu}\bot}\equiv{{\anti\pi}_{\ol{\mu\nu}\bot}\over J}
    \approx 0.
\end{eqnarray}
The non-zero PB's are
\begin{eqnarray}
   \{\ps\phi,{\symm\phi}{}_{\ol{\mu\nu}}^\prime\}
    &=&{4\over3}{\de_{xx^\prime}\over J^2}\e_{\lan\mu}{}^{\tau\s\bot}
    \tens\pi_{\ol\nu\ran\ol{\tau\s}},\label{s2ma}\\
   \{\phi_\bot,\phi^\prime\}
    &=&-{\de_{xx^\prime}\over J}{6a_0\over l^2},\label{s2mb}\\
   \{\phi_\bot,{\symm\phi}{}_{\ol{\mu\nu}}^\prime\}
    &=&{\de_{xx^\prime}\over J^2}\symm\pi_{\ol{\mu\nu}\bot},\label{s2mc}\\
   \{\vect\phi_{\ol\mu},\phi_{\ol\nu\bot}^\prime\}
    &=&{\de_{xx^\prime}\over J}\left[{1\over J}{\symm\pi}_{\ol{\mu\nu}\bot}
    -{2a_0\over l^2}\eta_{\ol{\mu\nu}}\right],\label{s2md}\\
   \{\vect\phi_{\ol\mu},{\anti\phi}{}_{\ol{\nu\s}}^\prime\}
    &=&{2\over3}{\de_{xx^\prime}\over J^2}\tens\pi_{\ol\mu[\ol{\nu\s}]},
    \label{s2me}\\
   \{\vect\phi_{\ol\mu},{\symm\phi}{}_{\ol{\nu\s}}^\prime\}
    &=&{\de_{xx^\prime}\over J^2}\tens\pi_{\ol{\s\mu\nu}},\label{s2mf}\\
   \{\anti\phi_{\ol{\mu\nu}\bot},\phi_{\ol\s\bot}^\prime\}
    &=&-{2\over3}{\de_{xx^\prime}\over J^2}\tens\pi_{\ol\s[\ol{\mu\nu}]},
    \label{s2mg}\\
   \{\anti\phi_{\ol{\mu\nu}\bot},{\anti\phi}{}_{\ol{\tau\s}}^\prime\}
    &=&{\de_{xx^\prime}\over J}\left[{1\over J}{\symm\pi}_{[\ol\tau[\ol\nu|\bot}
    \eta_{|\ol\mu]\ol\s]}+{a_0\over l^2}\eta_{\ol\s[\ol\mu}\eta_{\ol\nu]\ol\tau}
    \right],\label{s2mh}\\
   \{\anti\phi_{\ol{\mu\nu}\bot},{\symm\phi}{}_{\ol{\tau\s}}^\prime\}
    &=&{\de_{xx^\prime}\over J^2}{\symm\pi}_{(\ol\tau[\ol\nu|\bot}
    \eta_{|\ol\mu]\ol\s)},\label{s2mi}
\end{eqnarray}
and the primary Poisson matrix has the form
\begin{equation}
   \begin{array}{ccc|cccccccc|}
      \multicolumn{3}{c}{}&1&3&3&5&1&1&3&\multicolumn{1}{c}{3}\\
      \multicolumn{3}{c}{}&\ob{}&\ob{}&\ob{}&\ob{}&\ob{}&\ob{}&
       \ob{}&\multicolumn{1}{c}{\ob{}}\\
      \multicolumn{3}{c}{}&\phi&\phi_{\ol\mu\bot}&{\anti\phi}_{\ol{\mu\nu}}&
      {\symm\phi}_{\ol{\mu\nu}}&\phi_\bot&\ps\phi&{\anti\phi}_{\ol{\mu\nu}\bot}&
       \multicolumn{1}{c}{{\vect\phi}_{\ol\mu}}\\
      \multicolumn{10}{c}{}\\
      1&\{&\phi&&&&&\eta&\bc&\bc&\bc\\
      3&\{&\phi_{\ol\mu\bot}&&&&&\bc&\bc&\pi&{\eta+}\pi\\
      3&\{&{\anti\phi}_{\ol{\mu\nu}}&
      \multicolumn{4}{c}{\raisebox{2.2ex}[5pt]{$\bc$}}&\bc&\bc&{\eta+}\pi&\pi\\
      5&\{&{\symm\phi}_{\ol{\mu\nu}}&&&&&\pi&\pi&\pi&\pi\\
      1&\{&\phi_\bot&\eta&\bc&\bc&\pi&&&&\\
      1&\{&\ps\phi&\bc&\bc&\bc&\pi&&&&\\
      3&\{&{\anti\phi}_{\ol{\mu\nu}\bot}&\bc&\pi&{\eta+}\pi&\pi&
       \multicolumn{4}{c|}{\raisebox{2.2ex}[5pt]{$\bc$}}\\
      3&\{&{\vect\phi}_{\ol\mu}&\bc&{\eta+}\pi&\pi&\pi&&&&
   \end{array}
\label{ts2m}
\end{equation}
The corresponding super-Hamiltonian and the reduced constraints are
\begin{eqnarray}
   \h_\bot
    &=&{\k\over3b_1}{\tens\pi}_{\ol{\s\mu\nu}}({{\tens\pi}
    {}^{\ol{\s\mu\nu}}\over J}+{2b_1\over\k}{\tens R}{}^{\ol{\mu\nu\s}}{}_\bot)
    -{a_0\over 2l^2}J{\ul R}\nonumber\\
   &&\qquad\qquad +{\k\over2b_1}{\symm\pi}_{\ol{\mu\nu}\bot}
    ({{\symm\pi}{}^{\ol{\mu\nu}\bot}\over J}+{2b_1\over\k}{\ul R}
    {}^{\lan\ol{\mu\nu}\ran})-n^\mu\na_a\pi^a{}_\mu,\\
   \h_a
    &\lr&{2\over3}{{\tens\pi}_{\ol{\mu\nu\s}}\over J}{\ul R}^{[\ol{\nu\s}]}
    +{{\tens\pi}_{\ol{\s\mu\nu}}\over J}{\ul R}^{\lan\ol{\nu\s}\ran}
    +{{\symm\pi}{}^{\ol{\nu\s}}{}_\bot\over J}R_{\ol{\mu\nu\s}\bot}
    +{a_0\over l^2}R_{\ol\mu\bot}\approx0,\\
   \h_{\ol\mu\bot}
    &\lr&{a_0\over l^2}{\vect T}_{\ol\mu}+{{\symm\pi}_{\ol{\mu\nu}\bot}\over J}
    {\vect T}{}^{\ol\nu}+{4\over3}{{\tens\pi}_{[\ol{\mu\nu}]}
    {}^{\ol\s}\over J}\na_{\ol\s}n^\nu+\eta_{\ol{\mu\nu}}\na_{\ol\s}
    {{\symm\pi}{}^{\ol{\nu\s}\bot}\over J}\approx0,\\
   \h_{\ol{\mu\nu}}
    &\lr&{4\over3}\left[{{\tens\pi}_{\ol\s[\ol{\mu\nu}]}\over J}
    {\vect T}{}^{\ol\s}+\eta_{\ol\tau[\ol\mu}\eta_{\ol\nu]\ol\rho}\na_{\ol\s}
    {{\tens\pi}{}^{\ol{\s\tau\rho}}\over J}\right]\nonumber\\
   &&\qquad\qquad -{2\over J}{\symm\pi}{}^{\ol\s}{}_{[\ol\mu|\bot}\na_{\ol\s}
    n_{|\nu]}+{a_0\over l^2}T_{\ol{\mu\nu}\bot}\approx0.
\end{eqnarray}

Guided by our earlier experience, we readily realize that after its
linearization the remaining non-zero
PB's turn out to be Eq.~(\ref{s2mb},\ref{s2md},\ref{s2mh}) in which only the
zero-order (constant) terms exist.
Then the consistency conditions guarantee that the linearized Lagrange
multipliers, i.e., $u^\fl$, $u^{\fl\bot}$, $u^{\fl\ol\mu\bot}$,
${\vect u}{}^{\fl\ol\mu}$, ${\anti u}{}^{\fl\ol{\mu\nu}}$ and
${\anti u}{}^{\fl\ol{\mu\nu}\bot}$, are obtained.
The six PIC's, ${\ps\phi}{}^\fl$ and ${\symm\phi}{}_{\ol{\mu\nu}}^\fl$,
and the six SC's, ${\ps\chi}{}^\fl$ and ${\symm\chi}{}_{\ol{\mu\nu}}^\fl$,
which are derived from the consistency conditions of ${\ps\phi}{}^\fl$
and ${\symm\phi}{}_{\ol{\mu\nu}}^\fl$ are
second-class pairs, respectively.
These second-class pairs determine the multipliers ${\ps u}{}^\fl$ and
${\symm u}{}^{\fl\ol{\mu\nu}}$ and the algorithm of the linearized case
is terminated.
The degrees of freedom of the linearized case are counted as
${1\over2}(80-40-20[\mbox{PIC}]-6[\mbox{SC}])=7=5[2^-]+2$,
a massive spin-$2^-$ field in addition to the usual graviton.

On the other hand, the full spin-$2^-$ case does not follow this track.
Generically, $u^\bot$, ${\vect u}{}^{\ol\mu}$, ${\anti u}{}^{\ol{\mu\nu}\bot}$
can be determined.
Since $\ps\phi$ does not commute with ${\symm\phi}_{\ol{\mu\nu}}$,
the consistency condition of ${\symm\phi}_{\ol{\mu\nu}}$ can determine
$\ps u$ and give four secondary constraints $\chi_{[4]}$.
The consistency conditions of the connection primary constraints, $\phi_\bot$,
$\ps\phi$, $\anti\phi_{\ol{\mu\nu}}$ and $\vect\phi_{\ol\mu}$,
leave four of the tetrad Lagrange multipliers, $u$, $u^{\ol\mu\bot}$,
${\anti u}{}^{\ol{\mu\nu}}$ and ${\symm u}{}^{\ol{\mu\nu}}$ undetermined.
From the experience of the linearized case, we infer that
the consistency condition of the four $\chi_{[4]}$ will allow
the four undetermined tetrad Lagrange multipliers to be worked out and
thereby terminate the process.
The degrees of freedom would then count as
${1\over2}(80-40-20[\mbox{PIC}]-4[\mbox{SC}])=8=5[2^-]+1+2$.
It appears that one degree of freedom of the spin-$2^+$ mode is excited
as well as the degrees of freedom that we expected.
This is rather suspicious; a more careful analysis should be done to
clarify these details.  However, the main point is certain: once again
the nonlinear effect makes the structure of the full simple spin-$2^-$
case different from its linearized case.

The ill-behaviors in the cases above show that the existence of a
single higher-spin mode is virtually impossible.
Generally, in these cases there are extra degrees of freedom and the
phenomenon of constraint bifurcation (compared with their linearized theories).
Here the extra degrees of freedom represent ghosts, i.e.,
propagating negative energy, and the phenomenon of constraint bifurcation
leads to tachyons, i.e., faster-than-light propagation.
Therefore, nonlinearity makes the full theory qualitatively different from
its linearization and the theory fails important theoretical tests.

The constraints coming from the reductions of the super-momenta and the Lorentz
rotation parts do not seem to be much help in determining the existence of
the extra degrees of freedom and the constraint bifurcation.
This is because these constraints can be used
to eliminate some unphysical canonical variables instead of
unphysical canonical momenta.
But only the canonical momenta are involved in the phenomena in these simple
cases.

The reason we do not show the explicit forms of the SC's in these cases
is because they are very complicated and we can obtain a sufficient
understanding with only their linearized forms.
The detailed form of these SC's is certainly needed in order to
solve for the explicit form of the Lagrange
multipliers.  But the main point is that we know that this will be
possible since they will not commute with all the primary
constraints (in particular with their second-class pair
counterparts).
The complication of all the SC's in the full nonlinear theories comes from
the entanglements of the variables and the momenta.
These complications will not decrease the possibility of constraint bifurcation
for the same reasons that were stated in the last paragraph.

\subsection{Simple spin-$0^-$ $+$ spin-$2^-$ cases}
\label{rab3}
\noindent
Although the inconsistency of the simple spin-$2^-$ case with its linearization
renders the theory unsuccessful, the sign of the few unexpected degrees of
freedom appearing after including the nonlinear terms is notable.  It becomes
straightforward to recognize the situation from Table \ref{cpc}. In Table
\ref{cpc} only the
PIC's $\ps\phi$ and ${\tens\phi}_{\ol{\s\mu\nu}}$ do not have their ``natural''
second-class pair counterparts coming from the tetrad PIC's; their second-class
pair counterparts are the SC's derived from their own
consistency conditions, i.e., $\ps{\dot\phi}$ and ${\tens{\dot\phi}}_{\ol
{\s\mu\nu}}$. Thus the status of these two modes is unconcerned with the
 the tetrad parameter choice even in the full nonlinear case.
This feature can be used advantageously; it will help in constructing some
``more viable'' simple cases.  Here we show
two parameter choices which give propagating $\ps\pi$ and
${\tens\pi}_{\ol{\s\mu\nu}}$ modes generically, albeit with drawbacks
in the possible degeneracy.

\subsubsection{The negative $b_2$ case}
\noindent
According to Table \ref{cpc} the parameter $b_2$ appears in ${\ps\phi}(0^-)$,
${\anti\phi} _{\ol{\mu\nu}\bot}(1^+)$ and ${\tens\phi}_{\ol{\s\mu\nu}}(2^-)$.
Here we make the tetrad parameters vanish in order to suppress the spin-$1^+$
mode.  Therefore only the spin-$0^-$ and spin-$2^-$ modes propagate.
the specific parameter choice is
\begin{equation}
 \begin{array}{l}
   a_0\ne 0,\quad b_2<0,\\
   a_1=a_2=a_3=0,\\
   b_1=b_3=b_4=b_5=b_6=0.
 \end{array}
 \end{equation}
 This choice leads to the PIC's
 \begin{eqnarray}
  \phi\equiv{\pi\over J}\approx 0,\quad
  &&{\symm \phi }_{\ol {\mu \nu }}\equiv {{\symm\pi }
           _{\ol {\mu \nu }}\over J}\approx 0,\nonumber\\
\phi _{\ol \mu \bot }
  \equiv {\pi _{\ol \mu \bot }\over J}\approx 0,\quad
  &&{\anti \phi }_{\ol {\mu \nu }}\equiv{{\anti \pi }
           _{\ol {\mu \nu }}\over J}\approx 0,\nonumber\\
\phi_\bot\equiv{\pi_\bot\over J}-{3a_0\over l^2}\approx 0,\quad
  &&{\vect \phi}_{\ol \mu}\equiv{{\vect \pi}_{\ol \mu}\over J}\approx 0,
\nonumber\\
  {\symm \phi}_{\ol{\mu \nu}\bot}\equiv{{\symm \pi}
           _{\ol{\mu \nu}\bot}\over J}\approx 0.\quad&&
 \end{eqnarray}
 The non-zero PB's are then
 \begin{eqnarray}
 \{\phi_\bot,\phi^\prime\}&=&-{\de_{xx^\prime}\over J}{6a_0\over l^2},
 \label{nb2a}\\
 \{\phi_\bot,{\anti\phi}{}_{\ol{\mu\nu}}^\prime\}&=&
 {\de_{xx^\prime}\over J^2}\anti\pi_{\ol{\mu\nu}\bot},\\
 \{\vect\phi_{\ol\mu},\phi_{\ol\nu\bot}^\prime\}&=&
 -{\de_{xx^\prime}\over J}\left[{1\over J}{\anti\pi}_{\ol{\mu\nu}\bot}
 +{2a_0\over l^2} \eta_{\ol{\mu\nu}}\right],\label{nb2b}\\
 \{\vect\phi_{\ol\mu},{\anti\phi}{}_{\ol{\nu\s}}^\prime\}&=&
 {\de_{xx^\prime}\over J^2}\left[{1\over6}{\ps\pi}\e_{\mu\nu\s\bot}+{2\over3}
\tens\pi_{\ol\mu[\ol{\nu\s}]}\right],\\
 \{\vect\phi_{\ol\mu},{\symm\phi}{}_{\ol{\nu\s}}^\prime\}&=&
 {\de_{xx^\prime}\over J^2}\tens\pi_{\ol{\s\mu\nu}},\\
 \{\symm\phi_{\ol{\mu\nu}\bot},\phi_{\ol\s\bot}^\prime\}&=&
 -{\de_{xx^\prime}\over J^2}\tens\pi_{\ol{\nu\s\mu}},\\
 \{\symm\phi_{\ol{\mu\nu}\bot},{\anti\phi}{}_{\ol{\tau\s}}^\prime\}&=&
 {\de_{xx^\prime}\over J^2}{\anti\pi}_{[\ol\tau\lan\ol\mu|\bot}
\eta_{|\ol\nu\ran\ol\s]},\\
 \{\symm\phi_{\ol{\mu\nu}\bot},{\symm\phi}{}_{\ol{\tau\s}}^\prime\}&=&
 {\de_{xx^\prime}\over J}\left[{a_0\over l^2}\eta_{\ol\tau\lan\ol\mu}
\eta_{\ol\nu\ran\ol\s}-{1\over J}\eta_{(\ol\tau(\ol\mu}
{\anti\pi}_{\ol\nu)\ol\s)\bot}\right].\label{nb2c}
\end{eqnarray}
The primary Poisson matrix is
\begin{equation}
   \begin{array}{ccc|ccccccc|}
      \multicolumn{3}{c}{}&1&3&3&5&1&3&\multicolumn{1}{c}{5}\\
      \multicolumn{3}{c}{}&\ob{}&\ob{}&\ob{}&\ob{}&\ob{}&
       \ob{}&\multicolumn{1}{c}{\ob{}}\\
      \multicolumn{3}{c}{}&\phi&\phi_{\ol\mu\bot}&{\anti\phi}_{\ol{\mu\nu}}&
       {\symm\phi}_{\ol{\mu\nu}}&\phi_\bot&{\vect\phi}_{\ol\mu}&
       \multicolumn{1}{c}{{\symm\phi}_{\ol{\mu\nu}\bot}}\\
      \multicolumn{10}{c}{}\\
      1&\{&\phi&&&&&\eta&\bc&\bc\\
      3&\{&\phi_{\ol\mu\bot}&&&&&\bc&{\eta+}\pi&\pi\\
      3&\{&{\anti\phi}_{\ol{\mu\nu}}&
       \multicolumn{4}{c}{\raisebox{2.2ex}[5pt]{$\bc$}}&\pi&\pi&\pi\\
      5&\{&{\symm\phi}_{\ol{\mu\nu}}&&&&&\bc&\pi&{\eta+}\pi\\
      1&\{&\phi_\bot&\eta&\bc&\pi&\bc&&&\\
      3&\{&{\vect\phi}_{\ol\mu}&\bc&{\eta+}\pi&\pi&\pi&&\bc&\\
      3&\{&{\symm\phi}_{\ol{\mu\nu}\bot}&\bc&\pi&\pi&{\eta+}\pi&&&
   \end{array}
\end{equation}
The corresponding super-Hamiltonian and the reduced constraints are now
\begin{eqnarray}
 \h_\bot&=&{\k\over24b_2}\ps\pi({\ps\pi\over J}-{2b_2\over\k}\ps R_{\circ\bot})
 +{\k\over3b_2}\tens\pi_{\ol{\s\mu\nu}}({\tens\pi{}^{\ol{\s\mu\nu}}\over J}
 -{2b_2\over\k}\tens R{}^{\ol{\mu\nu\s}}{}_\bot)\nonumber\\
 &&+{\k\over2b_2}{\anti\pi}_{\ol{\mu\nu}\bot}({{\anti\pi}{}^{\ol{\mu\nu}\bot}
\over J}+{2b_2\over\k}{\ul R}{}^{[\ol{\mu\nu}]})
 -{a_0\over 2l^2}J{\ul R}-n^\mu\na_a\pi^a{}_\mu,\\
\h_a&\lr&\left[{1\over6}{\ps\pi\over J}\e_{\mu\nu\s\bot}
+{2\over3}{{\tens\pi}_{\ol{\mu\nu\s}}\over J}\right]{\ul R}^{[\ol{\nu\s}]}
+{{\tens\pi}_{\ol{\s\mu\nu}}\over J}{\ul R}^{\lan\ol{\nu\s}\ran}\nonumber\\
&&\qquad\qquad\qquad
+{{\anti\pi}{}^{\ol{\nu\s}}{}_\bot\over J}R_{\ol{\mu\nu\s}\bot}
+{a_0\over l^2}R_{\ol\mu\bot}\approx0,\\
\h_{\ol\mu\bot}&\lr&{a_0\over2l^2}{\vect T}_{\ol\mu}-{{\anti\pi}_{\ol{\mu\nu}
\bot}\over J}{\vect T}{}^{\ol\nu}+{1\over12}{\ps\pi\over J}\e_{\mu\nu\s\bot}
T^{\ol{\nu\s}}{}_\bot\nonumber\\
&&\qquad\qquad+{4\over3}{{\tens\pi}_{[\ol{\mu\nu}]}{}^{\ol\s}\over
J}\na_{\ol\s}n^\nu+\eta_{\ol{\mu\nu}}\na_{\ol\s}{{\anti\pi}{}^{\ol{\nu\s}\bot}
\over J}\approx0,\\
\h_{\ol{\mu\nu}}&\lr&
{4\over3}\left[{{\tens\pi}_{\ol\s[\ol{\mu\nu}]}\over J}{\vect T}{}^{\ol\s}
+\eta_{\ol\tau[\ol\mu}\eta_{\ol\nu]\ol\rho}\na_{\ol\s}
{{\tens\pi}{}^{\ol{\s\tau\rho}}\over J}\right]
-{2\over J}{\anti\pi}{}^{\ol\s}{}_{[\ol\mu|\bot}\na_{\ol\s}n_{|\nu]}\nonumber\\
&&\quad-{1\over6}\left[{\ps\pi\over J}\e_{\mu\nu\s\bot}{\vect T}{}^{\ol\s}
+{1\over6}\e^\s{}_{\mu\nu\bot}\na_{\ol\s}{\ps\pi\over J}\right]
+{a_0\over l^2}T_{\ol{\mu\nu}\bot}\approx0.
\end{eqnarray}
Although this case seems more complicated, we found that the
theory now has generically the same structure as its linearization.
Briefly
Eq.~(\ref{nb2a}, \ref{nb2b}, \ref{nb2c}) are the only non-vanishing PB's after
linearization
because of their constant terms. Then the linearized multipliers $u^\fl$,
$u^{\fl\bot}$,  $u^{\fl\ol\mu\bot}$, ${\vect u}{}^{\fl\ol\mu}$, ${\symm
u}{}^{\fl\ol{\mu\nu}}$ and
${\symm u}{}^{\fl\ol{\mu\nu}\bot}$ can be determined from the consistency
conditions of the linearized PIC's $\phi^\fl$, $\phi^\fl_\bot$,
$\phi_{\ol\mu\bot}^\fl$, ${\vect\phi}{}_{\ol
\mu}^\fl$, ${\symm\phi}{}_{\ol{\mu\nu}}^\fl$ and ${\symm\phi}{}_{\ol
{\mu\nu}\bot}^\fl$. Since ${\anti\phi}{}_{\ol{\mu\nu}}^\fl$
commutes with all the other linearized PIC's, its consistency condition gives
the linearized SC's, ${\anti\chi}{}_{\ol{\mu\nu}}^\fl$, i.e.,
\begin{equation}
{\rmd\over\rmd t}{\anti\phi}{}_{\ol{\mu\nu}}^\fl\approx0\lr
{\anti\chi}{}_{\ol{\mu\nu}}^\fl={{\anti\pi}{}_{\ol{\mu\nu}\bot}^\fl\over J^\fl}
-{2b_2\over\k}{\ul R}_{[\ol{\mu\nu}]}^\fl\approx0.
\end{equation}
As expected ${\anti\pi}{}_{\ol{\mu\nu}\bot}^\fl$  appears in
${\anti\chi}{}_{\ol{\mu\nu}}^\fl$, so
${\anti\chi}{}_{\ol{\mu\nu}}^\fl$ and ${\anti\phi}{}_{\ol{\mu\nu}}^\fl$ are a
second-class pair. Therefore the SC's can be used to determine the multiplier
${\anti u}{}^{\fl\ol{\mu\nu}}$. At the same time ${\anti\chi}{}_{\ol{\mu\nu}}
^\fl$ also suppresses the the spin-$1^+$ mode ${\anti\pi}{}_{\ol{\mu\nu}\bot}
^\fl$ and prevents the system from having the problem of a propagating negative
energy mode.

Generically the structure of the full theory is consistent with the one of the
linearized theory despite the fact that the analysis is not as straightforward
as the description in the last paragraph.  The key difference is that
${\anti\phi}_{\ol{\mu\nu}}$ does not commute with the connection PIC's
nonlinearly. The
consistency condition,
\begin{eqnarray}
{\rmd\over\rmd t}{\anti\phi}_{\ol{\mu\nu}}&=&\int\{{\anti\phi}_{\ol{\mu\nu}},
\h_\rmT^\prime\}\nonumber\\
&\approx&\int\left[\{{\anti\phi}_{\ol{\mu\nu}},N^\prime\h_\bot^\prime\}
+u^{\bot\prime}\{{\anti\phi}_{\ol{\mu\nu}},\phi_\bot^\prime\}
\right.\nonumber\\
&&\qquad\qquad
\left.+{\vect u}{}^{\ol\s\prime}\{{\anti\phi}_{\ol{\mu\nu}},
{\vect\phi}{}_{\ol\s}^\prime\}
+{\symm u}{}^{\ol{\mu\nu}\bot\prime}
\{{\anti\phi}_{\ol{\mu\nu}},{\symm\phi}{}_{\ol{\tau\s}\bot}^\prime\}\right]
\approx0,\label{nb2d}
\end{eqnarray}
can give the SC ${\anti\chi}_{\ol{\mu\nu}}$,
here the connection multipliers
shown in Eq.~(\ref{nb2d}) can be derived from the consistency conditions of the
other tetrad PICs. ${\anti\chi}
_{\ol{\mu\nu}}$ includes ${\anti\chi}{}_{\ol{\mu\nu}}^\fl$ as the linear part
and thus associates with ${\anti\phi}_{\ol{\mu\nu}}$ to become
a second-class pair. Consequently, the SC renders ${\anti\phi}_{\ol{\mu\nu}
\bot}$ non-dynamical. Now we have enough constraints to determine all the
multipliers and finish the process. The degrees of freedom are
${1\over2}(80-40-21[\mbox{PIC}]-3[\mbox{SC}]) =8=1[0^-]+5[2^-]+2$
in both the linear and nonlinear
theory. The result indicates that the theory can be transformed into its weak
field approximation without qualitative changes in the canonical
structure.

However the theory cannot entirely escape from the constraint bifurcation
problem. By
using {\tt REDUCE} the possible vanishing of the determinant of the sub-Poisson
matrix formed by the PB of $\phi_{\ol\mu\bot}$, ${\symm\phi}_{\ol{\mu\nu}}$,
${\vect\phi}_{\ol\mu}$ and ${\symm\phi}_{\ol{\mu\nu}\bot}$ is affirmed, i.e.,
with certain numerical values of the canonical variables which satisfy
all constraints the sub-Poisson matrix can be singular.
This is not so good.  Such an occurrence is thought to be a sign of the
presence of a mode which propagates faster than light\cite{Chen98}.

\subsubsection{The $b_1+b_3$ case}
\noindent
The other parameter choice available to activate only the spin-$0^-$ and
spin-$2^-$ modes
is to let all parameters vanish except $a_0$, $b_1$ and $b_3$. This choice gives
the PIC's $\phi$, $\phi_{\ol\mu\bot}$, ${\anti\phi}_{\ol{\mu\nu}}$,
${\symm\phi}_{\ol{\mu\nu}}$, $\phi_\bot$, ${\anti\phi}_{\ol{\mu\nu}\bot}$ and
${\vect\phi}_{\ol\mu}$. The non-zero PB's are
 \begin{eqnarray}
 \{\phi_\bot,\phi^\prime\}&=&-{\de_{xx^\prime}\over J}{6a_0\over l^2},\\
 \{\phi_\bot,{\symm\phi}{}_{\ol{\mu\nu}}^\prime\}&=&
 {\de_{xx^\prime}\over J^2}{\symm\pi}_{\ol{\mu\nu}\bot},\\
 \{{\anti\phi}_{\ol{\mu\nu}\bot},\phi_{\ol\s\bot}^\prime\}&=&
 {\de_{xx^\prime}\over J^2}\left[{2\over3}{\tens\pi}_{[\ol{\mu\nu}]\ol\s}
 -{1\over6}{\ps\pi}\e_{\mu\nu\s\bot}\right],\\
 \{{\anti\phi}_{\ol{\mu\nu}\bot},{\anti\phi}{}_{\ol{\tau\s}}^\prime\}&=&
 {\de_{xx^\prime}\over J}\left[{a_0\over l^2}\eta_{\ol\s[\ol\mu}
 \eta_{\ol\nu]\ol\tau}+{1\over J}{\symm\pi}_{[\ol\s[\ol\mu|\bot}
 \eta_{|\ol\nu]\ol\tau]}\right],\\
 \{{\anti\phi}_{\ol{\mu\nu}\bot},{\symm\phi}{}_{\ol{\tau\s}}^\prime\}&=&
 {\de_{xx^\prime}\over J^2}
\eta_{[\ol\mu(\ol\s}{\symm\pi}_{\ol\tau)\ol\nu]\bot},\\
 \{\vect\phi_{\ol\mu},\phi_{\ol\nu\bot}^\prime\}&=&
 {\de_{xx^\prime}\over J}\left[{1\over J}{\symm\pi}_{\ol{\mu\nu}\bot}
 -{2a_0\over l^2} \eta_{\ol{\mu\nu}}\right],\\
 \{\vect\phi_{\ol\mu},{\anti\phi}{}_{\ol{\nu\s}}^\prime\}&=&
 {\de_{xx^\prime}\over J^2}\left[{1\over6}{\ps\pi}\e_{\mu\nu\s\bot}+{2\over3}
\tens\pi_{\ol\mu[\ol{\nu\s}]}\right],\\
 \{\vect\phi_{\ol\mu},{\symm\phi}{}_{\ol{\nu\s}}^\prime\}&=&
 {\de_{xx^\prime}\over J^2}\tens\pi_{\ol{\s\mu\nu}}.
\end{eqnarray}
and the primary Poisson matrix is
\begin{equation}
   \begin{array}{ccc|ccccccc|}
      \multicolumn{3}{c}{}&1&3&3&5&1&3&\multicolumn{1}{c}{3}\\
      \multicolumn{3}{c}{}&\ob{}&\ob{}&\ob{}&\ob{}&\ob{}&
       \ob{}&\multicolumn{1}{c}{\ob{}}\\
      \multicolumn{3}{c}{}&\phi&\phi_{\ol\mu\bot}&{\anti\phi}_{\ol{\mu\nu}}&
       {\symm\phi}_{\ol{\mu\nu}}&\phi_\bot&{\anti\phi}_{\ol{\mu\nu}\bot}&
       \multicolumn{1}{c}{{\vect\phi}_{\ol\mu}}\\
      \multicolumn{10}{c}{}\\
      1&\{&\phi&&&&&\eta&\bc&\bc\\
      3&\{&\phi_{\ol\mu\bot}&&&&&\bc&\pi&{\eta+}\pi\\
      3&\{&{\anti\phi}_{\ol{\mu\nu}}&
      \multicolumn{4}{c}{\raisebox{2.2ex}[5pt]{$\bc$}}&\bc&{\eta+}\pi&\pi\\
      5&\{&{\symm\phi}_{\ol{\mu\nu}}&&&&&\pi&\pi&\pi\\
      1&\{&\phi_\bot&\eta&\bc&\bc&\pi&&&\\
      3&\{&{\anti\phi}_{\ol{\mu\nu}\bot}&\bc&\pi&{\eta+}\pi&\pi&&\bc&\\
      3&\{&{\vect\phi}_{\ol\mu}&\bc&{\eta+}\pi&\pi&\pi&&&
   \end{array}
\end{equation}
The corresponding super-Hamiltonian and the reduced constraints are
\begin{eqnarray}
 \h_\bot&=&{\k\over24b_3}{\ps\pi}({\ps\pi\over J}+{2b_3\over\k}
 {\ps R}_{\circ\bot})
 +{\k\over3b_1}{\tens\pi}_{\ol{\s\mu\nu}}({{\tens\pi}{}^{\ol{\s\mu\nu}}\over J}
 +{2b_1\over\k}{\tens R}{}^{\ol{\mu\nu\s}}{}_\bot)\nonumber\\
 &&+{\k\over2b_1}{\symm\pi}_{\ol{\mu\nu}\bot}({{\symm\pi}{}^{\ol{\mu\nu}\bot}
\over J}+{2b_1\over\k}{\ul R}{}^{\lan\ol{\mu\nu}\ran})
 -{a_0\over 2l^2}J{\ul R}-n^\mu\na_a\pi^a{}_\mu,
\end{eqnarray}
\begin{eqnarray}
\h_a&\lr&\left[{1\over6}{\ps\pi\over J}\e_{\mu\nu\s\bot}
+{2\over3}{{\tens\pi}_{\ol{\mu\nu\s}}\over J}\right]{\ul R}^{[\ol{\nu\s}]}
+{{\tens\pi}_{\ol{\s\mu\nu}}\over J}{\ul R}^{\lan\ol{\nu\s}\ran}\nonumber\\
&&\qquad\qquad\qquad
+{{\symm\pi}{}^{\ol{\nu\s}}{}_\bot\over J}R_{\ol{\mu\nu\s}\bot}
+{a_0\over l^2}R_{\ol\mu\bot}\approx0,\\
\h_{\ol\mu\bot}&\lr&{a_0\over l^2}{\vect T}_{\ol\mu}+{{\symm\pi}_{\ol{\mu\nu}
\bot}\over J}{\vect T}{}^{\ol\nu}+{1\over12}{\ps\pi\over J}\e_{\mu\nu\s\bot}
T^{\ol{\nu\s}}{}_\bot\nonumber\\
&&\qquad\qquad+{4\over3}{{\tens\pi}_{[\ol{\mu\nu}]}{}^{\ol\s}\over
J}\na_{\ol\s}n^\nu+\eta_{\ol{\mu\nu}}\na_{\ol\s}{{\symm\pi}{}^{\ol{\nu\s}\bot}
\over J}\approx0,\\
\h_{\ol{\mu\nu}}&\lr&
{4\over3}\left[{{\tens\pi}_{\ol\s[\ol{\mu\nu}]}\over J}{\vect T}{}^{\ol\s}
+\eta_{\ol\tau[\ol\mu}\eta_{\ol\nu]\ol\rho}\na_{\ol\s}
{{\tens\pi}{}^{\ol{\s\tau\rho}}\over J}\right]
-{2\over J}{\symm\pi}{}^{\ol\s}{}_{[\ol\mu|\bot}\na_{\ol\s}n_{|\nu]}\nonumber\\
&&\quad-{1\over6}\e^\s{}_{\mu\nu\bot}
\left[{\ps\pi\over J}{\vect T}_{\ol\s}+\na_{\ol\s}{\ps\pi\over J}\right]
+{a_0\over l^2}T_{\ol{\mu\nu}\bot}\approx0.
\end{eqnarray}

Essentially the same argument as used in the negative $b_2$ case can be applied
to this case.  The degrees of freedom are generically
${1\over2}(80-40-19[\mbox{PIC}]-5[\mbox{SC}])=8=1[0^-]+5[2^-]+2$ in both the
linear and nonlinear theory.
But once again it is readily seen that this case also
almost certainly has constraint bifurcation and consequently is very
vulnerable to acausal propagation.

Two questions are raised by these results:  (a) Could
relaxed
versions of the two spin-$0^-$ + $2^-$ modes overcome the
 defect of constraint
bifurcation?  (b) Do there exist any parameter choices which make
 the compound
propagating modes nonlinearly viable?  Concerning the first question, according
to the explanation in the beginning of this chapter, more nonlinear
effects will
be involved in the relaxed versions.  The Poisson matrix is supposed to be more
complicated and even include the spatial derivatives of $\de_{xx^\prime}$.  It
will become  harder to analyze the whole system and to convince people that it
has
the advantage of being free of any constraint bifurcation problems.
 To the second question,
basically this is beyond our ability to give a definite answer.  Maybe with some
delicate tuning of all parameters there exist such nonlinearly viable cases
of the PGT.
(After all, unexpected miraculous cancellations have been found in other
systems.)
However, based on our experience with the PGT, we
hold a very conservative attitude to the possibility of this happening.

In order to convince ourselves that the cases dubbed ``simple'' in this
section do show the obstacles that all the relaxed cases and
 many similar cases
could meet in the Hamiltonian analysis,
it seems worthwhile to investigate the most
promising parameter choice coming from the linearized PGT.
In section 4 
the parameter choice (\ref{rab1}) given by Kuhfuss
and Nitsch\cite{KRNJ86} was quite restrictive and supposed to be ``viable''.
Here we give a brief analysis of the theory with this parameter choice.
Referring to Table \ref{cpc} we learn that the PIC's in the theory are
$\phi_{\ol\mu\bot}$, ${\anti\phi}_{\ol{\mu\nu}}$, $\phi_\bot$,
${\anti\phi}_{\ol{\mu\nu}\bot}$ and ${\symm\phi}_{\ol{\mu\nu}\bot}$.
Under linearization one will expect that ${\anti\phi}{}_{\ol{\mu\nu}}^\fl$ and
${\anti\phi}{}_{\ol{\mu\nu}\bot}^\fl$ are first-class due to $a_1=a_0$.
And $\phi_{\ol\mu\bot}^\fl$, $\phi_\bot^\fl$ and
${\symm\phi}{}_{\ol{\mu\nu}\bot}^\fl$ commute with one another.
Subsequently $\chi_{\ol\mu\bot}^\fl$, $\chi_\bot^\fl$ and
${\symm\chi}{}_{\ol{\mu\nu}\bot}^\fl$ are derived from their parent constraints.
These SC's and their parent constraints form second-class pairs and thus
suppress $\pi_{\ol\mu\bot}^\fl$, $\pi_\bot^\fl$ and
${\symm\pi}{}_{\ol{\mu\nu}\bot}^\fl$ respectively.
The degrees of freedom count as
${1\over2}(80-40-15[\mbox{PIC}]-9[\mbox{SC}]-6[\mbox{gauges}])=5
=2[\mbox{GR}]+3$.
It is easy to discover that the three degrees of freedom come from
the massless spin-$0^-$ and spin-$2^-$ propagating modes.

Now we return to the full nonlinear case.
In order to make the analysis simpler, we can specify the parameters to be
$b_1=b_4=b_6=0$.
First, $\phi_{\ol\mu\bot}$ and ${\anti\phi}_{\ol{\mu\nu}\bot}$ will not
commute with the connection PIC's because of the nonlinear terms.
Consequently we expect that the consistency conditions of all the PIC's
will determine twelve Lagrange multipliers and give three
 SC's $\chi_{[3]}$.
Using $\chi_{[3]}$, the remaining three multipliers can be obtained.
The degrees of freedom would then count as
${1\over2}(80-40-15[\mbox{PIC}]-3[\mbox{SC}])=11=2[\mbox{GR}]+1[0^-]
+3[1^-]+5[2^-]$; all excited propagating modes have become ``massive''
formally.  Nonlinear effects have excited extra dynamic degrees of freedom
and destroyed the gauge freedom of the linearized theory.
Moreover, if
 we require the kinetic energy density of the spin-$2^-$ mode to be
positive definite, it is inevitable that the spin-$1^-$ mode propagates with
negative energy density.  This unexpected mode with the wrong sign shows
another defect of this parameter choice.

Here we try to illustrate that the relaxed cases can be even more
 vulnerable to
getting stuck in troubles.
Essentially, the Kuhfuss \& Nitsch's case is in the same track as ours in
Section 5.2. 
They concluded that only the spin-$0^-$ and spin-$2^-$ modes could propagate.
We found from the nonlinear structure of the PGT that it is usually
unlikely to suppress one single higher-spin ($s>0$) mode without suppressing
its counterpart if they form a second-class pair in the linearized theory.
This realization forces the parameter choice to be the one which only
activates the modes not belonging to any second-class pairs ---
the spin-$0^-$ and spin-$2^-$ modes.

Maybe there exists still the thought that it might still work with
different values of the nonzero parameters or in more relaxed versions.
Until we do further more detailed investigations this possibility cannot
be dismissed. Along this line one has to face more problems:
(1) Because the connection PIC's do not commute generally in the relaxed
cases, more non-zero PB's appear, which could excite extra degrees of
 freedom.
Negative energy can be expected to  appear in most of those cases;
(2) In general, the spatial derivatives of the $\de_{xx^\prime}$ function
could appear in the PB's of the connection PIC's.
This greatly complicates the difficulty of analyzing the generic rank of
the Poisson matrix.
Even if these problems are  settled down well there is the very possible
phenomenon of constraint bifurcation with its attendant acausal modes
waiting ahead!

\section{Discussion and Conclusion}
\noindent
In this paper we examined the behavior of a restricted case of the MAG ---
the PGT, the local gauge theory of the Poincar\'e group.
People used to work on the linearized PGT and its initial value problems to
clarify its viability, however they overlooked the significant influences
of nonlinearity on the whole theory.
In  Sec.~3 
we noted that two problems are
produced by nonlinearity:
(1) {\it Constraint bifurcation}:  the phenomenon is caused by the appearances
of the nonlinear terms in the constraints which lead to a field dependence
in the Poisson matrix.
This can cause a bifurcation in the constraint chain.
The number and type of constraints can depend on the field values.
This phenomenon has been linked to acausal (tachyonic) propagation
modes\cite{HRNZ94,Chen98}.
(2) {\it Field activation}: the phenomenon is that nonlinearity turns
some original constraints in the linearized theories into field equations.
This means that some field modes which should be frozen are activated.
If the fields carry negative energy, it violates the
 ``no-ghost''    requirement.
We regard these two phenomena as very important;
 our criteria is that they should be avoided.
A good theory should keep the same dynamical structure before and after
linearization; it should not be bothered by these two phenomena.
We find it hard to imagine the consequences of a theory of
a fundamental
interaction which would exhibit a different dynamical structure as we
passed from the strong field to weak field regime.

In the last section we examined nonlinear effects in the PGT which can
switch the status of presumably viable parameter choices.
The main tools are the {\it Dirac-Bergmann algorithm} of Hamiltonian analysis,
and the {\it if-constraint} technique developed by Blagojevi\'c and Nikoli\'c
(described in Sec.~3). 
The procedures we used are as follows:
\begin{itemize}
\item[(a)] Make the desired parameter choice according to the study of the
linearized PGT.
\item[(b)] Identify all the constraints in the selected case and calculate
their Poisson brackets.
\item[(c)] Classify the constraints by the results of the Poisson brackets.
From the consistency conditions find out all secondary constraints and
the Lagrange multipliers.
\item[(d)] Count the degrees of freedom of the selected case.
By knowing the corresponding linearized results, we can figure out
the meaning of each degree of freedom without further calculation.
\item[(e)] Evaluate whether the Poisson matrix formed by the Poisson brackets
of all the constraints can be singular at specific values of the variables
in order to determine any occurrence of constraint bifurcation.
The computer algebraic software {\tt REDUCE} can be applied to work out
the lengthy calculations.
\end{itemize}

Individual ``simple'' spin modes, which are well behaved and thus viable in
the linearized PGT, were chosen to study their behavior under the full
nonlinear considerations.
For the spin zero modes, we previously found that they are essentially
viable although nonlinear effects obviously complicate the whole systems
We noted that there is the possibility of constraint bifurcation, however,
this can be avoided with specific value choices of the
parameters\cite{YHNJ99}.
Here we concluded that the nonlinear terms simply devastate the
 viabilities of the spin one modes
and spin two modes, since each expectedly frozen mode which has the same
spin but opposite parity with the propagating mode in every case is activated
by nonlinearity.
Unfortunately, the nonlinear-activated mode propagates with negative energy
which is physically unacceptable theoretically.

There can be maximally three different propagating torsion modes in the
linearized PGT as described in Sec.~4. 
Therefore we looked for the possibly viable multi-modes according to
the classification in Ref.~7. 
Through many trial-and-errors only two ``simple'' cases of multi-modes which
circumvent the difficulty of the activation of the unexpected modes are found,
i.e., keep only the spin-$0^-$ and spin-$2^-$ modes alive and suppress
all the other spin modes.
Successfully the structure (and the degrees of freedom) of the full nonlinear
theory remains the same as that of its linearization in these two case.
This is a promising results. Even though we argued
that the constraint bifurcation of the Poisson matrices could be
expected to render them unattractive, we wish to note that that issue, and
these modes more generally, are worthy of a deeper study.

There are a few other special parameter choices meriting examiniation,
 e.g., those identified by Katanaev which give solutions with vanishing
curvature or torsion\cite{KatM93}.
  One that we specifically considered here was
Kuhfuss and Nitsch's case, which was thought to be
the most promising case for being viable.  We found that the case changes
its character drastically between the nonlinear and linearized versions.
Both the phenomena of constraint bifurcation and field activation happen;
they devastate the case.
Considering this result, one could be pessimistic about finding viable
higher spin modes in the nonlinear PGT.
But we don't exclude the possibility, with a fine tuning of all the
involved parameters.
Once the case exists, the propagating modes are expected to be short-range,
i.e., massive, just like the viable spin zero modes.

It seems that only a few cases in the PGT can avoid the phenomenon of
field activation. And almost all of these cases are plagued with the
constraint bifurcation problem.
This means that the nonlinear PGT is very likely to be qualitatively
different from the linearized PGT in the number and type of constraints.
The more general MAG theory includes the PGT as a subcase and thus
already has these same problems.  Moreover it seems extremely likely that
nonmetricity itself will be vulnerable to the same kind of problems.

Our analysis of non-linear constraint effects in the PGT
reveals an extreme difficulty in finding a viable gauge
alternative of gravity which had not been appreciated in the past. In the
light of this, a further important consequence of our analysis is
an enhanced understanding of the advantages and uniqueness of GR.

\section*{Acknowledgments}
\noindent
This work was supported by the National Science Council of the R.O.C.
under grants No. NSC90007P and NSC90-2112-M-008-041.

\section*{References}
\noindent

\end{document}